# Patterns of protection, infection, and detection:

# Country-level effectiveness of COVID 19 vaccination in reducing mortality worldwide

Cosima Rughiniș, Department of Sociology, University of Bucharest, Bucharest, 030167, Romania, cosima.rughinis@unibuc.ro

Mihai Dima, Department of Physics, University of Bucharest, Bucharest, 030167, Romania, mihai.dima@unibuc.ro

Simona-Nicoleta Vulpe, Interdisciplinary School of Doctoral Studies, University of Bucharest, Bucharest, 050107, Romania, simona.vulpe@drd.unibuc.ro

Răzvan Rughiniș, Department of Computer Science and Engineering, University POLITEHNICA of Bucharest, Bucharest, 060042, Romania, razvan.rughinis@cs.pub.ro

Sorina Vasile, Interdisciplinary School of Doctoral Studies, University of Bucharest, Bucharest, 050107, Romania, sorina.vasile@gmail.com

**Corresponding author:**

Simona-Nicoleta Vulpe, 36-46 Mihail Kogălniceanu, room 221 first floor, Sector 5, Bucharest, 050107, Romania. E-mail: simona.vulpe@drd.unibuc.ro

# Patterns of protection, infection, and detection: Country-level effectiveness of COVID-19 vaccination in reducing mortality worldwide

**Abstract**

We investigated the negative relationship between mortality and COVID-19 vaccination at ecological level, which has been established through clinical trials and other investigations at the individual level. We conducted an exploratory, correlational, country-level analysis of open data centralized by Our World in Data concerning the cumulative COVID-19 mortality for the winter wave (October 2021–March 2022) of the pandemic as function of the vaccination rate in October 2021. At country level, patterns of vaccine protection have not been clearly differentiated from patterns of COVID-19 infection and detection. In order to disentangle the protective relationship from confounding processes, we controlled variables that capture country-level social development and level of testing. We also deployed three segmentation tactics, distinguishing among countries based on their level of COVID-19 testing, age structure, and types of vaccines used. Controlling for confounding factors did not highlight a statistically significant global relationship between vaccination and cumulative mortality in the total country sample. As suggested by previous estimates at country level, a strong, significant, negative relationship between cumulative mortality (log scale) and vaccination was highlighted through segmentation analysis for countries positioned at the higher end of the social development spectrum. The strongest estimate for vaccine effectiveness at ecological level was obtained for countries that use Western-only vaccines. This may partly reflect the higher effectiveness of Western vaccines in comparison with the average of all vaccines in use; it may also derive from the lower social heterogeneity of countries included in this segment, which minimizes confounding influences. COVID-19 testing (log scale) has a significant and positive relationship with cumulative mortality for all subsamples, consistent with patterns of under- and overreporting of COVID-19 deaths at country level, partly driven by testing. This indicates that testing intensity should be controlled as a potential confounder in future ecological analyses of COVID-19 mortality.

**Keywords**: COVID-19 vaccine; vaccine effectiveness; mortality; COVID-19 testing; ecological study.

# 1 Introduction

The efficacy and effectiveness of COVID-19 vaccines are documented in multiple studies at the individual and country levels. Still, ecological estimates of vaccine effectiveness against COVID-19 mortality face the challenge of being disentangled from the confounding factors that have shaped the pandemic's impact. Specifically, the COVID-19 pandemic has been more intense in countries with higher income and social development [1, 2], which are, at the same time, the countries with higher vaccination rates.

It has been proven, through stage III clinical trials and other individual-level studies, that COVID-19 vaccines significantly reduce the risk of dying for people infected with COVID-19 [2, 3]. Vaccine protection persists across successive variants [4]. Ecological analyses also reported negative correlations between COVID-19 vaccination and mortality at the country level. Some of these studies focused on highly developed countries, such as those in Europe and North America, and Israel, thus circumventing the positive Human Development Index (HDI) mortality correlation, which is specifically for wider global samples, but not for the subsamples of high-income countries [5]. A study of 32 countries in Europe and Israel found a high effectiveness of vaccination against death, through a time series analysis of new COVID-19 deaths from January 2020 through April 2021 [6]. An investigation of 30 countries of the European Economic Area (EEA) until January 2022 reported that low vaccination rates, high proportions of older people, low funding, and inadequate staffing of public health systems were independent risk factors for a higher case fatality rate [7]. A study of the 27 European Union (EU) countries found that countries with higher vaccine coverage improved their relative cumulative mortality profile within the EU [8]. Another study of 50 countries belonging to the World Health Organization (WHO) European Region, in addition to the USA and Canada, indicated that vaccination coverage was strongly and negatively associated with excess mortality [9].

When broadening the geographical focus, a study of 184 countries from December 2020 to December 2021 found that an increased vaccination rate significantly suppressed new deaths per million, and that a threshold of 70% additionally contributed to protection from death through herd immunity [10]. Longitudinal research covering 90 countries in the interval from November 2020 to April 2021 concluded that "an increase in one vaccinated person per 10 people in the population, or a 10% increase in the vaccine coverage, reduced the CFR [COVID-19 case fatality ratio] by approximately 7.6%" [11]. A study of the global relationship between vaccination rates and COVID-19 outcomes up to August 2021 did not find proof of vaccination effectiveness in longitudinal data, but observed that cross-sectionally, at a given time point, and especially for the subset of countries with high vaccination rates (more than 60%), there was a negative association between vaccination rates and new deaths per million [12].

Estimating the global impact of vaccination on COVID-19 mortality is challenging because many countries with low vaccination levels have also reported reduced rates of infections, case fatality, and mortality. For example, the *relatively low intensity of the pandemic in the African continent, despite very low vaccination rates,* remains a topic of scientific and public debate, being attributed to factors such as a young population, higher proportion of rural populations with lower density and more outdoor living, lower international travel, cross-immunity from previous infections, climate factors, early and drastic lockdown and restriction policies, and possible underreporting, partly derived from low testing rates [14, 15, 13].

Thus, the published literature has identified at least three important factors that link development and vaccination with mortality in diverging directions [1] (Fig. 1). Two factors account for the higher COVID-19 mortality in more developed countries, while the third accounts for a negative influence of development on mortality. The roles played by these factors in the pandemic are discussed below.

First, as regards the testing level, there may be *differential methods of detection – that is, measurement and attribution*. Countries with fewer resources available have used fewer tests to diagnose COVID-19 infections, and testing is positively associated with HDI and with government capacity [14]. Thus, for countries with lower testing levels, some of the deaths that would have been otherwise identified and attributed to COVID-19 might have been attributed to other causes, or might have even remained unreported [15]. Reported COVID-19 deaths might derive from different tactics to differentiate those who die *with* COVID-19 from those who die *from* COVID-19, and widespread testing could make a difference in such tactics [16].

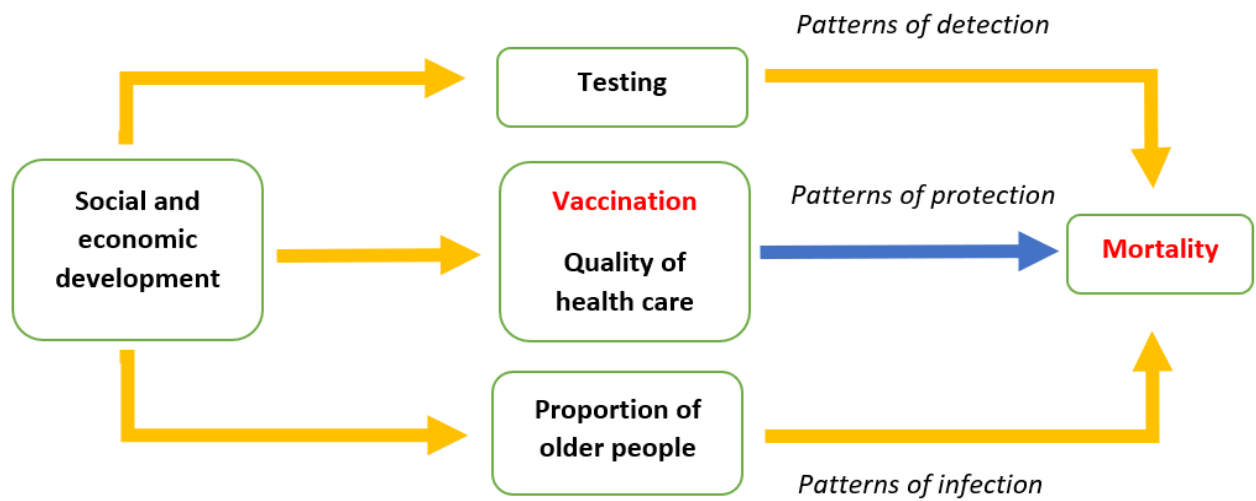

**Fig. 1.** Confounding processes for detecting the association of COVID-19 vaccination with mortality at the ecological level. Positive relationships are colored orange; negative relationships are colored blue. Source: Authors' representation.

Relatively few studies have examined COVID-19 testing as a proxy for country-level measurement strategies concerning deaths, and *testing is not, as a rule, controlled as a potential confounder for ecological analyses of COVID-19 mortality*. Several studies documented that early-stage pandemic testing capacity was associated with decreased mortality, as testing allowed better diagnosis [17, 18] and contact tracing. In this case, testing is an indicator of governmental pandemic response effectiveness. Analyses from Our World in Data (OWID) document linear positive associations both between total tests per million (log scale) and total confirmed COVID-19 deaths per million (log scale)[19], and between new tests per million (log scale) with new cases per million (log scale) [20, 21]. A study of member countries of the Organization for Economic Co-operation and Development (OECD), BRIC nations (Brazil, Russia, India, and China), and Taiwan reported a weak, positive, significant linear correlation between total deaths and total tests [22], although the low level of association may reflect the fact that variables were used in linear rather than log scale.

As regards the second factor, *countries with higher social and economic development have specific patterns of infection because of larger proportions of older people*, who have been particularly at risk in the COVID-19 pandemic [23]. Therefore, the proportion of older people in the general population may also account for the positive correlation between COVID-19 mortality and social and economic development metrics [24].

As regards the third factor, a *negative association between social development and mortality may be mediated by patterns of protection, through COVID-19 vaccination and the quality of health care*. Richer and more developed countries have had increased economic and political access to buy vaccines and have been allocated more resources for vaccination campaigns. Since the inception of the vaccination campaign, there has been a strong positive association between vaccination rates and the social and economic development of countries [10], which is symptomatic of high vaccine inequity. Thus, while social development is negatively linked with COVID-19 mortality through level of vaccination and access to effective health care, it is also positively linked with mortality through COVID-19 testing and attribution of deaths, and through higher proportions of older people.

In this study, we aimed to disentangle divergent factors that have shaped COVID-19 mortality and provide estimates for the negative association of vaccination with mortality at the ecological level by examining the relationship between vaccination rates in October 2021 and cumulative global mortality for the 2021–2022 winter COVID-19 wave. For this purpose, we combined visual explorations of bivariate associations with multiple linear regression models, and we explored and evaluated three strategies for segmenting the global country population in relevant subsamples.

## 2    Data and methods

We used publicly available data from OWID [25] to estimate the effectiveness of COVID-19 in mitigating cumulative global mortality in the winter of 2021–2022 wave of the novel coronavirus pandemic, covering October 2021 to March 2022.

We focused on the winter wave because it allowed us to better capture the impact of vaccination through a cross-sectional comparison of heterogeneous countries, leaving aside cumulative mortality accumulated in the first and second waves when, for many countries, vaccination campaigns were still incipient. For example, a study of the 27 EU countries found that vaccination impact on cumulative mortality only started to become visible with a 4-month lag after July 2021 [8]. As a visual illustration, in Supplementary Material 1, we have centralized waves in terms of cases and deaths for the 33 countries that only used Western vaccines, a category that we studied in one of our segmentation analyses described below.

In our analysis, we aimed to cover a wide diversity of countries across the globe and to disambiguate vaccination protection from other confounding factors, including the influence of testing on mortality measurements, the influence of societal age structures on mortality, and other imprints of social development (see Fig. 1), such as the reduction of mortality through quality of health care. We attempted to control for the latter by including in our analysis the dimensions of HDI created by the United Nations Development Program (UNDP) [26]. For each country, HDI aggregates three dimensions of development: economic prosperity as measured by gross national income (GNI) per capita, a long and healthy life as captured by life expectancy at birth, and human resources for development as measured by a combination of mean and expected years of schooling. Because the proportion of people aged 65 and over captures COVID-19 mortality risks better than does life expectancy, we used this variable instead of life expectancy, with which it is strongly correlated ($R$ = 0.752, *P-value* = 0.000).

Therefore, we included six variables at country level in the analysis, for which descriptive statistics are available in Table S.M. 1 in Supplementary Material 2. The dependent variable is *cumulative mortality for a 6-month period*, obtained as the difference between average cumulative mortality for the month of March 2022 and average cumulative mortality for the month of October 2021 (source: OWID). The main independent variable is the *vaccination rate*, representing the share of fully vaccinated people (%), averaged across the month of October 2021 (source: OWID). We also controlled for four confounding factors. *Total tests per thousand* represents a metric that aggregates available testing information from a subset of countries globally [20] (source: OWID, data reported for the first week of January 2022). The *proportion of people aged 65 or more per hundred*, *GNI per capita*, and *mean years of schooling* are reported by UNDP in its dataset for the Human Development Report [26], and they are also available via the OWID dataset on COVID-19 [25].

We included countries with a population larger than one million in the analysis, subject to the availability of data in the OWID dataset. The list of the 136 countries included in the analysis and the three types of segmentation are presented in Tables S.M. 5, S.M.6, and S.M.7 in Supplementary Material 2.

## 3 Results

An initial exploration of the global correlation patterns between the cumulative mortality in the winter wave of the COVID-19 pandemic (October 2021–March 2022) and the vaccination rate led to rather inconclusive results (see Chart 1.1 in Fig. 2). As discussed above, the dispersed scatterplot in Chart 1.2 of Fig. 2 likely resulted from overlapping the negative influence of vaccination on mortality with the positive relationships between social development and COVID-19 mortality.

Since mortality rates do not seem to be linearly related to either vaccination rates or HDI, we opted to use a *logarithmic scale for cumulative mortality*. This transformation led to a clearer visualization of the relationships (see the corresponding Charts 1.3 and 1.4). As shown in Chart 1.3, the cumulative mortality in the winter wave (log scale) had a nonlinear, reverse J-shaped pattern of association with vaccination rates: an initial positive relationship is followed by a plateau extending to the right, and then by an apparently negative association at the higher vaccination end of the spectrum. The positive linear relationship between cumulative mortality (log scale) and HDI is clearly visible in Chart 1.4.

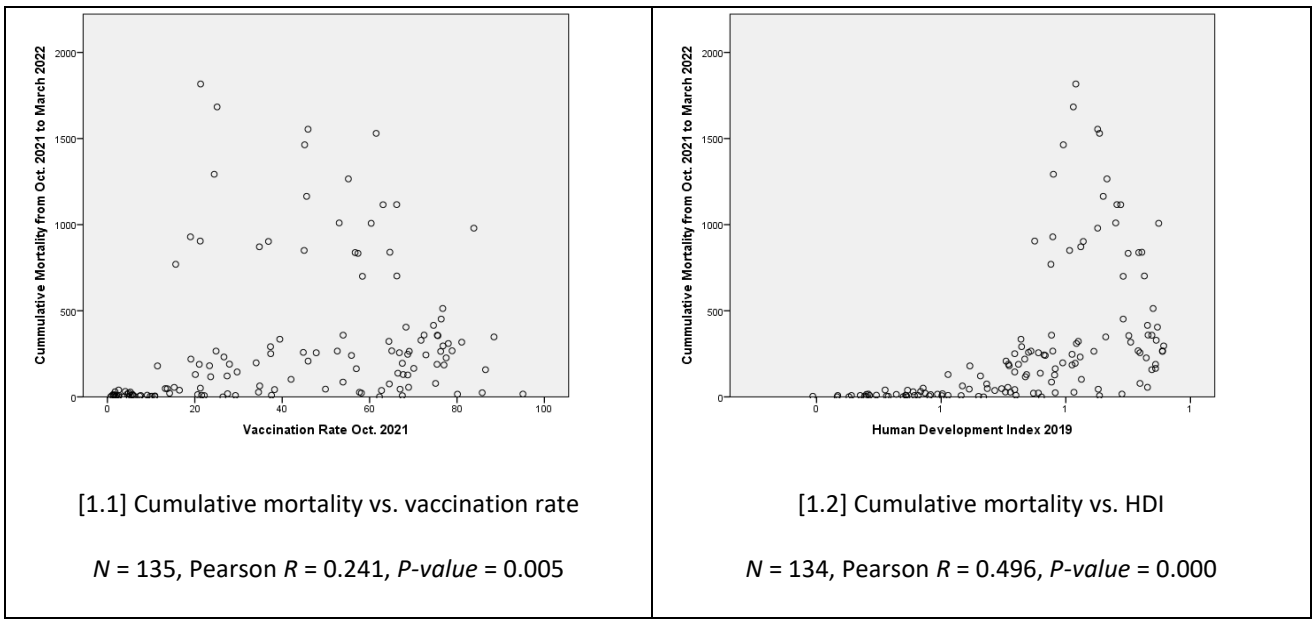

[1.1] Cumulative mortality vs. vaccination rate

*N* = 135, Pearson *R* = 0.241, *P-value* = 0.005

[1.2] Cumulative mortality vs. HDI

*N* = 134, Pearson *R* = 0.496, *P-value* = 0.000

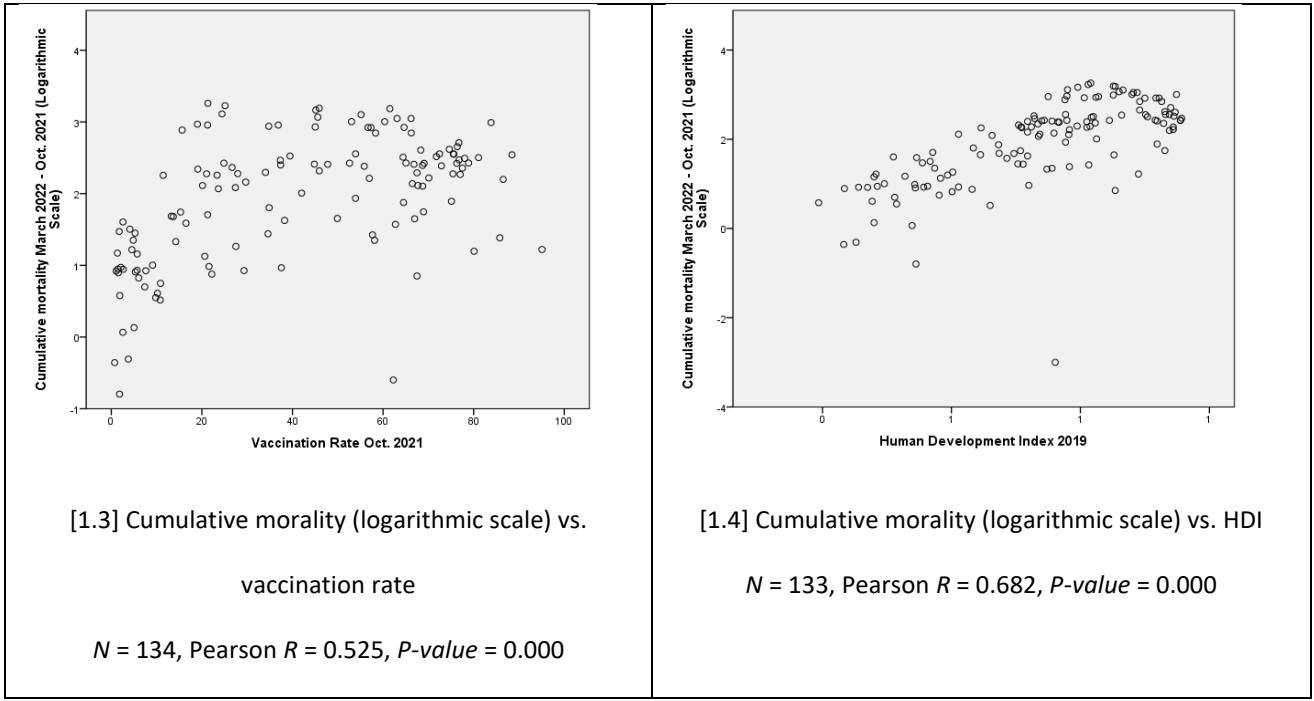

[1.3] Cumulative morality (logarithmic scale) vs. vaccination rate

*N* = 134, Pearson *R* = 0.525, *P-value* = 0.000

[1.4] Cumulative morality (logarithmic scale) vs. HDI

*N* = 133, Pearson *R* = 0.682, *P-value* = 0.000

**Fig. 2.** Exploration of covariation patterns between COVID-19 cumulative mortality and both the vaccination rate and the HDI

The positive association between social development and COVID-19 cumulative mortality in the 2021–2022 winter wave of the pandemic was possibly mediated by pandemic measurement strategies and by age structures. These two relationships, together with the associations with GNI per capita and with mean years of study at national level, are investigated and represented in Fig. 3.

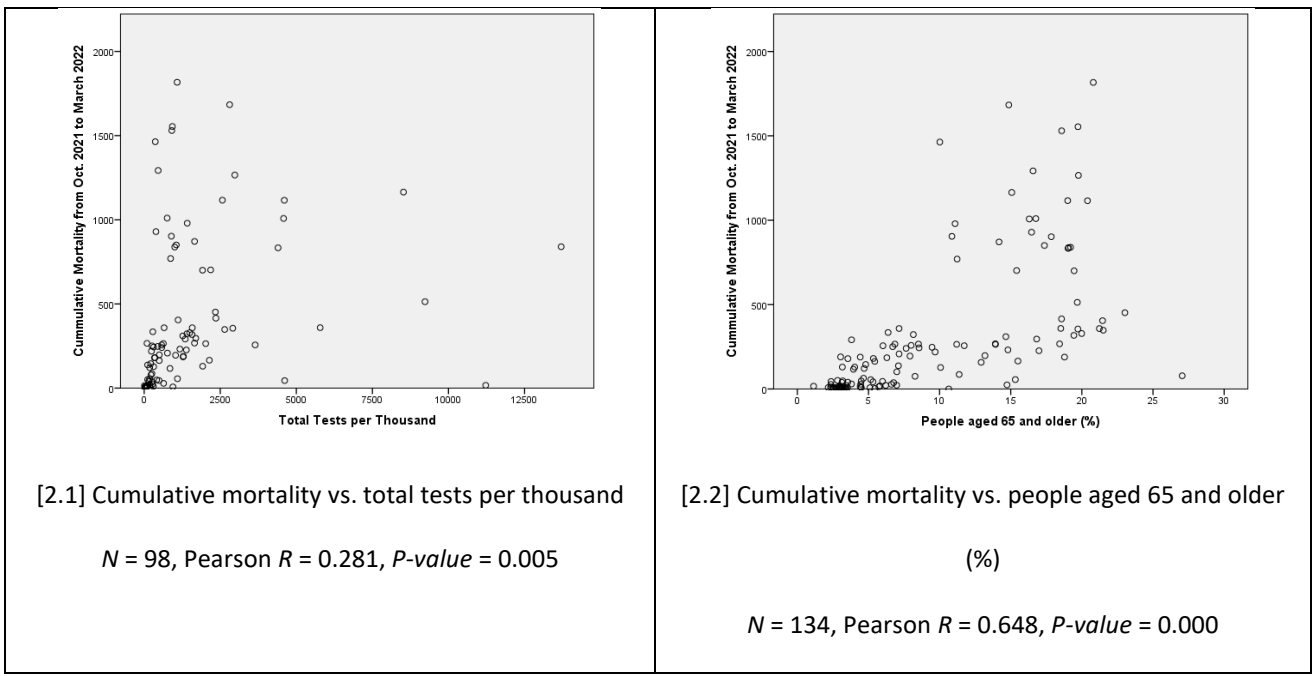

[2.1] Cumulative mortality vs. total tests per thousand

*N* = 98, Pearson *R* = 0.281, *P-value* = 0.005

[2.2] Cumulative mortality vs. people aged 65 and older (%)

*N* = 134, Pearson *R* = 0.648, *P-value* = 0.000

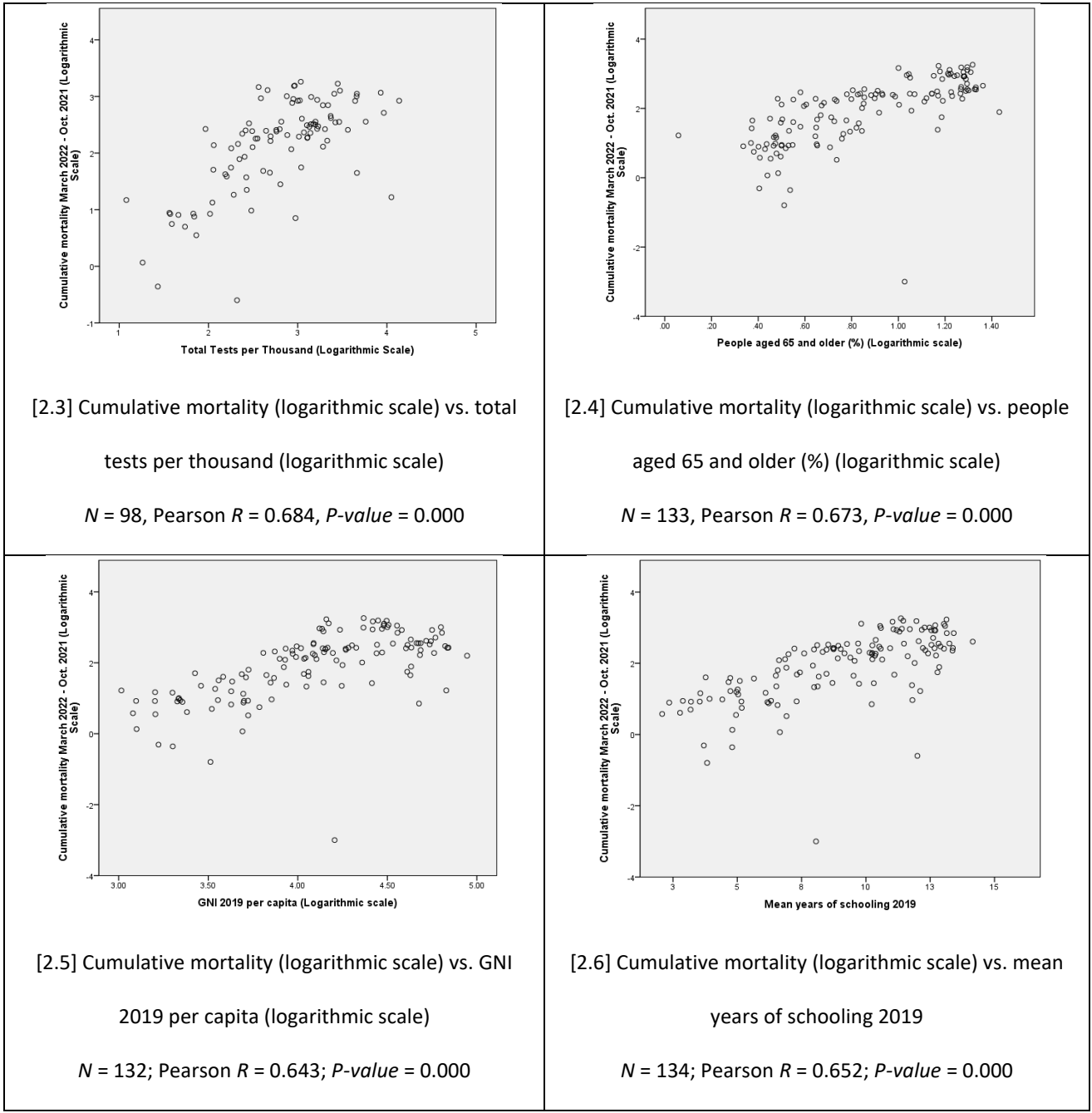

**Fig. 3.** Exploration of covariation patterns between COVID-19 cumulative mortality and the testing rate, the proportion of people aged 65 and older, GNI per capita, and mean years of schooling with linear vs. logarithmic scaling for all countries.

Based on previous findings, we expected that highly developed countries would have tested their populations more extensively, thus diagnosing more cases and possibly attributing more deaths to COVID-19 than countries with few resources allocated for testing. However, the scatterplot of cumulative mortality and tests per thousand (Chart 2.1) does not clarify the association because the relationship is not linear. Transforming both variables through a logarithmic scale shows a clear positive correlation between them (Chart 2.2). Similarly, the nonlinear positive association of cumulative mortality with the proportion of people aged 65 and older in the population (Chart 2.3) is clarified when both variables are visualized through a logarithmic transformation (Chart 2.4). Charts 2.5 and 2.6 present the linear associations of cumulative mortality (log scale) with GNI per capita (log scale) and mean years of schooling (linear scale), respectively.

## 3.1 Global analysis: Linear regression – All countries

In order to disentangle patterns of vaccine protection from divergent factors that have shaped COVID-19 infection and detection, we used a linear regression model to control for possible confounding variables. Table 1 presents the regression model for all countries included in the analysis. The model has a high predictive value with an adjusted $R^2$ of 70%. The only statistically significant coefficients are those that lead to the positive relationships between mortality and development, namely the influence of detection (testing), infection risk factors (proportion of older people), and other sources of positive covariation captured by country-level human capital (mean years of schooling). In the overall cross-sectional picture of the 2021–2022 winter wave of the COVID-19 pandemic, economic capital is not a statistically significant predictor of mortality when controlling for the other factors, nor is vaccination rate.

**Table 1**
Regression model for cumulative mortality from October 2021 to March 2022 (logarithmic scale) as a function of the vaccination rate in October 2021, controlling for measurement intensity, age structure, education structure, and GNI per capita.

|  | Unstandardized Coefficients | | Standardized Coefficients | | |
|---|---|---|---|---|---|
|  | B | SE | Beta | t | P-value |
| (Constant) | .548 | .808 |  | .679 | .499 |
| Vaccination rate Oct. 2021 | −.004 | .003 | −.117 | −1.221 | .225 |
| Total tests per thousand (logarithmic scale) | **.483** | .129 | **.400** | 3.732 | .000 |
| People aged 65 and older (%) (logarithmic scale) | **1.240** | .206 | **.514** | 6.025 | .000 |
| Mean years of schooling 2019 | **.087** | .036 | **.317** | 2.447 | .016 |
| GNI 2019 per capita (logarithmic scale) | −.373 | .281 | −.210 | −1.327 | .188 |

Dependent variable: Cumulative mortality for March 2022–October 2021 (logarithmic scale). $N = 90$, adjusted $R^2 = 0.703$.

## 3.2 Segmentation analysis

The positive but plateauing shape of the bivariate relationship between cumulative mortality and vaccination, as visualized in Chart 1.3 in Fig. 2, suggests the possibility of segmenting the population of countries into subcategories, attempting to better isolate the negative relationship between vaccination and mortality from other influences that country-level properties have on mortality.

Generally speaking, segmentation rests on the assumption that the global pattern conflates distinct and possibly divergent patterns that could be revealed by separating the population into relevant segments. If the global relationship is reversed through segmentation, the situation becomes an instance of the so-called Simpson's paradox (or the amalgamation paradox) [27]. It is important to keep in mind that any segmentation may actually work as a proxy for a different confounding variable that differentiates countries, leading to yet another version of the pervasive confounding risks of correlational and ecological analyses. With this caveat, we discuss in the following section the three segmentation tactics we used, which allowed us to distinguish countries through their pandemic coping strategies: by testing level, through their age structure, and by vaccine type. The three segmentation criteria were correlated, though not overlapping (see Table S.M. 11 in Supplementary Material 2).

We first segmented the country set through their approach to testing, comparing the subsample of countries with testing values below the median (758.85 tests per thousand) with those tested at or above median values. This segmentation was based on the assumption that higher testing is conducive to more precise measurement of COVID-19 mortality, and thus it is better suited to indicate the protective impact of vaccination at the country level.

We then segmented the global set of countries according to their proportion of people aged 65 and older, comparing the subsample of countries with values below the median (6.92%) with countries at or above the median. This segmentation worked on the assumption that the impact of vaccination on cumulative mortality is higher or better visible at the ecological level for countries with older populations, which had higher risks of COVID-19 deaths and thus benefitted more from vaccination, compared with countries with younger populations.

Finally, we segmented the country set according to the vaccines they used, comparing countries that administered non-Western vaccines, either exclusively or in combination with Western vaccines, with countries that only administered Western vaccines (Pfizer – BioNTech – Comirnaty, Moderna – Spikevax, Johnson & Johnson's – Janssen, and Oxford AstraZeneca – Vaxzevria).

A synthesis of segmentation results is presented in Table 2, indicating for the global sample and each segment, the partial correlation between cumulative mortality for the winter COVID-19 wave (log scale) and vaccination rate (October 2021), as well as controlling for the other variables in the model. *The segments with values above the median for the differentiating criterion and the Western-only vaccine segment attest negative, statistically significant associations between the cumulative COVID-19 mortality and the vaccination rate, when confounding factors are controlled*. No significant relationships were identified for the global, unsegmented sample or for the other segments. The strongest relationship was determined to be that of the Western-only vaccine set of countries, which was also the most exclusive in terms of membership (listwise *N* = 25 because of missing values for testing, and listwise *N* = 27 when testing is not controlled).

**Table 2**
Partial correlation between cumulative mortality (log scale) and vaccination rate, controlling for proportion of people aged 65 or more (log scale), tests per thousand (log scale), GNI per capita (log scale), and mean years of schooling.

| **Model** | **Segments** | **Also controlling for testing intensity** | | | **Without controlling for testing intensity** | | |
|---|---|---|---|---|---|---|---|
| | | Listwise *N* | Partial correlation | *P*-value | Listwise *N* | Partial correlation | *P*-value |
| Global analysis | All countries | 90 | –0.128 | 0.225 | 125 | –0.124 | 0.165 |
| Segmented by testing intensity | Testing at or above median | 42 | **–0.304** | 0.045 | 43 | –0.226 | 0.135 |
| | Testing below median | 42 | –0.027 | 0.862 | 43 | 0.049 | 0.747 |
| | Missing information for testing | N/A | | | 29 | –0.128 | 0.493 |
| Segmented by age structure | People aged 65+ at or above median | 51 | **–0.355** | 0.09 | 60 | **–0.522** | 0.000 |
| | People aged 65+ below median | 33 | –0.186 | 0.284 | 60 | 0.038 | 0.770 |
| Segmented by vaccine type | Western-only vaccines | 25 | **–0.616** | 0.001 | 27 | **–0.527** | 0.003 |
| | All vaccines | 59 | –0.013 | 0.918 | 93 | –0.073 | 0.484 |

If we do not control for testing level, we gain more cases in the analysis. The broad pattern of results is unchanged, with the exception of the partial correlation in the segment with testing above the median, which is no longer statistically significant. The negative correlation of vaccination rate with cumulative mortality (log scale) is still visible for the Western-only vaccine segment and for the segment of countries with older populations.

As shown in Fig. 4, we began by visually exploring the *segmentation of countries based on the distribution of the testing level*. We compared countries testing below the median with countries with testing above the median. The list of countries in each segment is available in Table S.M 4 in Supplementary Material 2. As Fig. 4 indicates, the countries with testing below the median displayed a positive, non-linear relationship between mortality (log scale) and vaccination, with a plateau toward higher values of vaccination (Chart 4.1), while countries with testing above the median displayed a negative, linear relationship between mortality (log scale) and vaccination (Chart 4.2).

We found a similar visual pattern for our second *segmentation based on the proportion of people aged 65 and older*, as shown in the second row of Fig. 4 (see the list of countries in Table S.M 6, Supplementary Material 2). Countries with a lower proportion of older people presented a positive, plateauing relationship between mortality (log scale) and vaccination, while countries with a higher proportion of older people presented a negative, linear relationship (see Charts 4.3 and 4.4, respectively).

Last but not least, as shown in the third row of Fig.4, Charts 4.5 vs. 4.6, we compared countries that have used only Western vaccines with countries that have used non-Western vaccines or a combination of the two. Countries belonging to each type are listed in Table S.M. 8, in Supplementary Material 2. The segment that used Western-only vaccines included 33 highly-developed countries, which had political and financial access to these vaccines and which, given their high social and economic development, also benefitted from better health care systems and testing policies, while running the mortality risks of an older population. This segmentation isolated a subsample of countries with a strong negative association between mortality and vaccination rates (the Western vaccines only segment, Chart 4.6) from a larger subsample that incorporated high heterogeneity and in which the negative association of vaccination with mortality was not visible at the ecological level (Chart 4.5). For the larger subsample of countries that also used non-Western vaccines, there was a positive bivariate correlation of vaccination and mortality because of the confounding association between social development and COVID-19 mortality.

The usefulness of this segmentation in revealing a subsample of countries with a strong, negative association between vaccination rate and mortality may partially be due to a higher effectiveness of Western vaccines in relation to non-Western vaccines for the winter 2021–2022 wave of COVID-19. Such a difference is compatible with results from some studies at the individual level [3, 4] and with possible differences in effectiveness, at the country level, between inactivated virus vaccines (such as Sinovac, Sinopharm, and Bharat Biotech, which are non-Western vaccines) and other types of vaccines (viral vector or genetic vaccines) [33]. However, our data cannot prove, disprove, or qualify such a difference because of the confounding processes at the ecological level that link vaccination rates and mortality rates to social development, age structures, testing, and measurement strategies, and other country characteristics.

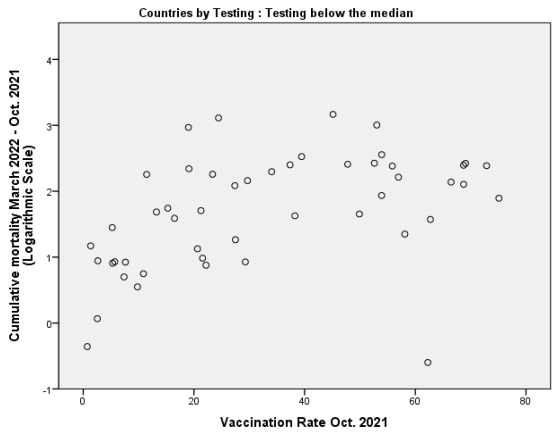

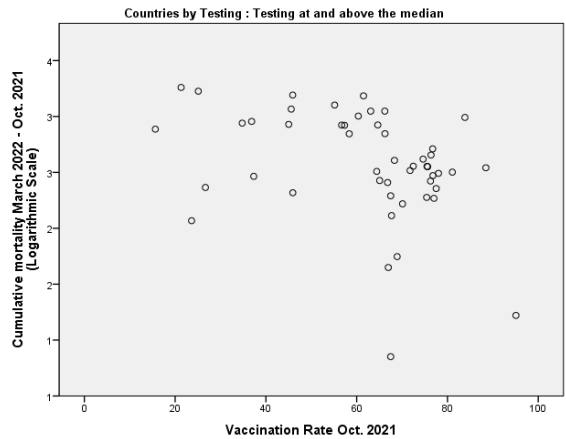

[4.1] Countries with **testing** rates below the median: Cumulative mortality vs. vaccination rate

$N$ = 49, Pearson $R$ = 0.457, $P$-value = 0.005

[4.2] Countries with **testing** rates at or above the median: Cumulative mortality vs. vaccination rate

$N$ = 49, Pearson $R$ = –0.371, $P$-value = 0.009

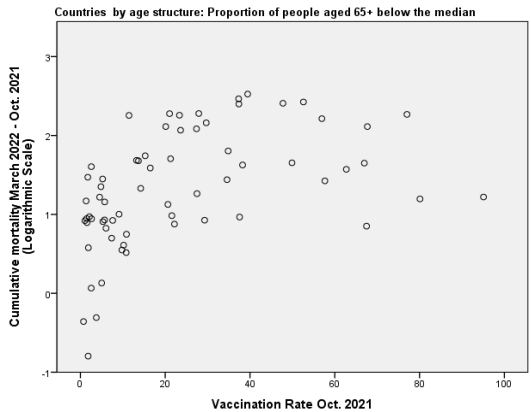

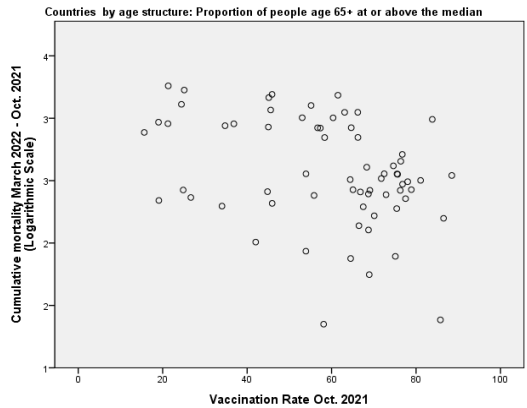

[4.3] Countries with people aged 65+ (%) below the median: Cumulative mortality vs. vaccination rate

$N$ = 66, Pearson $R$ = 0.467, $P$-value = 0.000

[4.4] Countries with people aged 65+ (%) at or above the median: Cumulative mortality vs. vaccination rate

$N$ = 66, Pearson $R$ = –0.366, $P$-value = 0.003

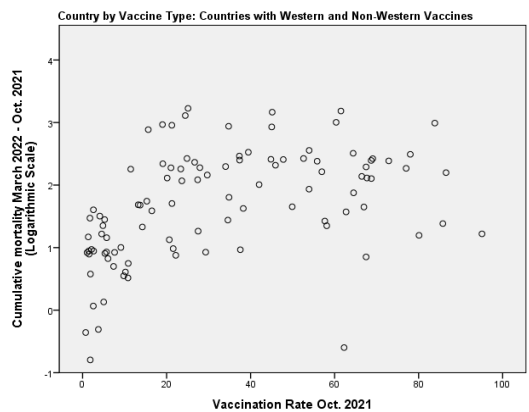

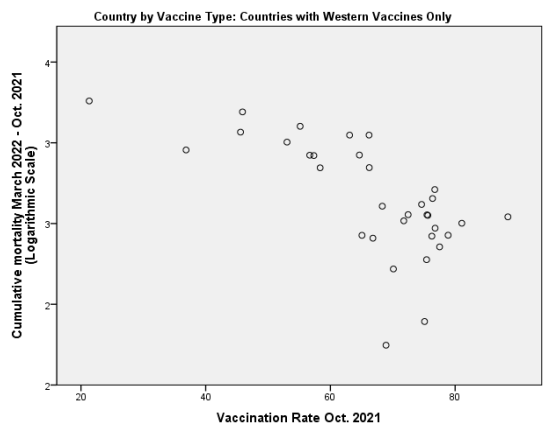

| [4.5] Countries with both **Western** and **Non-Western** vaccines: Cumulative mortality vs. vaccination rate  $N$ = 101, Pearson $R$ = 0.445, *P-value* = 0.000 | [4.6] Countries with **Western** vaccines only: Cumulative mortality vs. vaccination rate  $N$ = 33; Pearson $R$ = –0.671, *P-value* = 0.000 |
|---|---|

**Fig. 4.** Exploring patterns of covariation between cumulative mortality and vaccination rate in three segmentation scenarios: volume of testing, age structure, and vaccine type.

In Table 3, we synthesize regression models for segmentation analysis based on testing level, age structure, and vaccine type. The three regression models are presented in detail in Supplementary Material 2, in Tables S.M. 11, S.M.12, and S.M.13, respectively.

**Table 3**
Regression model for segmentation analysis based on testing intensity, age structure, and vaccine type.

| Segmentation: | Countries by testing | | | | Countries by proportion of people aged 65+ | | | | Countries by vaccine type | | | |
|---|---|---|---|---|---|---|---|---|---|---|---|---|
| **Countries:** | Below the median | | At or above the median | | Below the median | | At or above the median | | All vaccines | | Western-only vaccines | |
| **Model properties:** | $N$ = 42 Adjusted $R^2$ = 0.711 | | $N$ = 42 Adjusted $R^2$ = 0.692 | | $N$ = 33 Adjusted $R^2$ = 0.611 | | $N$ = 51 Adjusted $R^2$ = 0.413 | | $N$ = 59 Adjusted $R^2$ = 0.752 | | $N$ = 25 Adjusted $R^2$ = 0.598 | |
| **Model variables:** | **Beta** | *P-value* | **Beta** | *P-value* | **Beta** | *P-value* | **Beta** | *P-value* | **Beta** | *P-value* | **Beta** | *P-value* |
| Vaccination rate Oct. 2021 | –.021 | .862 | **–.244** | .045 | –.194 | .284 | **–.415** | .009 | –.010 | .918 | **–.683** | .001 |
| Total tests per thousand (logarithmic scale) | **.408** | .001 | **.282** | .004 | **.479** | .024 | **.478** | .001 | **.261** | .031 | **.454** | .001 |
| People aged 65 and older (%) | .163 | .259 | **.808** | .000 | **.514** | .000 | .352 | .024 | **.515** | .000 | .232 | .084 |
| (logarithmic scale) | | | | | | | | | | | | |
| Mean years of schooling 2019 | **.365** | .018 | .009 | .940 | **.498** | .009 | .006 | .971 | **.346** | .006 | −.074 | .623 |
| GNI 2019 per capita (logarithmic scale) | .047 | .788 | **−.346** | .018 | −.172 | .420 | −.218 | .357 | −.086 | .539 | −.082 | .679 |

Remarkably, we found that *the level of testing has a positive and statistically significant relationship with mortality in all analyzed segments, and it is the only predictor with such degree of consistency*. The proportion of older people has a strong, positive association with mortality in countries with testing above the median, in countries with older populations below the median, and in countries that used non-Western vaccines. We also found that GNI had a broadly negative association with mortality, but it was only statistically significant for countries with testing above the median. In contrast, the mean years of schooling had a strong and positive association with COVID-19 cumulative mortality in the winter wave for the three segments in which vaccination was not a significant predictor, thus capturing the pattern of positive correlations between development and mortality for these subsets of countries.

## 4 Conclusions

In this study, we attempted *to investigate at the ecological level the negative relationship between COVID-19 mortality and vaccination*, previously established through clinical trials and other investigations at the individual level. However, at the country level, *patterns of vaccine protection are obfuscated through patterns of COVID-19 infection and detection*. In order to disentangle the protective relationship from confounding processes, we controlled variables that captured the level of testing and country-level social development. Besides controlling for confounding factors, we also deployed segmentation tactics, distinguishing between countries on the basis of three criteria: level of COVID-19 testing, age structure, and type of vaccines in use. We thus conducted an exploratory, correlational, country-level analysis of open data centralized by OWID concerning the cumulative COVID-19 mortality for the winter wave (October 2021–March 2022) of the pandemic as a function of vaccination rate at Oct. 2021, total tests per thousand in early January 2022, and dimensions of social development, namely, the proportion of people aged 65 and more, GNI per capita, and mean years of schooling.

*Controlling for confounding factors did not highlight a statistically significant relationship between cumulative mortality for the winter wave and vaccination*, at the ecological level, globally (see Table 1). The linear

regression model estimated for all countries had a high predictive value (Adjusted $R^2$ = 70%), but the significant predictors were those that exclusively modelled the positive relationship between social development and COVID-19 mortality, at the aggregate level, worldwide. The level of testing, the proportion of older people, and the mean years of schooling had positive, significant associations with cumulative mortality in the winter 2021–2022 wave of COVID-19.

As suggested by previous estimates at the country level, a *strong, significant, negative relationship between cumulative mortality and vaccination was highlighted through segmentation analysis for countries positioned at the higher end of the social development spectrum* (see Table 2). When controlling for testing and social development indicators, the partial correlation of cumulative mortality (log scale) and vaccination was $R = -0.304$ for countries above the median of testing intensity ($N = 42$), $R = -0.355$ for countries with a proportion of people aged 65 and more above the median ($N = 51$), and $R = -0.616$ for countries that only used Western vaccines ($N = 25$).

Linear regression models for segmented subsets of countries highlight the fact that *COVID-19 testing (log scale) had a significant and positive relationship with cumulative mortality for all segments*. This finding is consistent with the published literature discussing possible patterns of under- and overreporting of COVID-19 deaths, partly driven by testing [16, 15]. Nevertheless, with the exception of some studies that have explored the link between early pandemic testing and reductions in mortality, there has been little use of updated testing information in multivariate models on COVID-19 mortality at the country level. This indicates that *testing intensity should be controlled as a potential confounder in future ecological analyses of COVID-19 mortality*.

*The stronger estimate for vaccine effectiveness at the ecological level for countries that used Western-only vaccines may reflect a higher effectiveness of these vaccines* in comparison with the average of all vaccines in use, *but it may also derive from the lower social heterogeneity of countries included in this segment*, which minimized other confounding influences. These countries have high social development levels, and they are predominantly located in the European Union and North America.

*Our study shares the limitations of large-scale, ecological, and correlational analyses*. The high number of countries included in analysis and their socio-political diversity means that measurements may include multiple sources of heterogeneity. Despite our goal to control for confounding factors for the relationship between vaccination and cumulative mortality, other sources of amalgamation may persist, obfuscating the relationship at the aggregate level. This limitation is especially salient for our analysis at the global level, in which we did not find a statistically significant relationship between vaccination and mortality, despite the good predictive power of the regression model. Finally, the impact of vaccination on mortality refers to processes that take place at the individual and interpersonal levels, such as preventing transmission, serious illness, and death. Therefore, a study at the country level would run the additional risk of ecological errors caused by successive steps of aggregation in the measurement processes.

## 5　Author contribution

All authors made a significant contribution to the development of this manuscript and approved the final version for submission.

## 6　Funding

This research was funded by the Executive Unit for Financing Higher Education, Research, Development and Innovation (UEFISCDI), Romania, grant PN-III-P4-ID-PCE-2020-1589 "The manufacture of doubt on vaccination and climate change. A comparative study of legitimacy tactics in two science-skeptical discourses," https://skepsis-project.ro/.

## 7　References

[1]　Shahbazi F, Khazaei S. Socio-economic inequality in global incidence and mortality rates from coronavirus disease 2019: an ecological study. New Microbes and New Infections 2020;38:100762. https://doi.org/10.1016/J.NMNI.2020.100762.

[2]　Mohammed I, Nauman A, Paul P, Ganesan S, Chen KH, Jalil SMS, et al. The efficacy and effectiveness of the COVID-19 vaccines in reducing infection, severity, hospitalization, and mortality: a systematic review. Hum Vaccin Immunother 2022;18. https://doi.org/10.1080/21645515.2022.2027160.

[3]　Mathieu E, Roser M. How do death rates from COVID-19 differ between people who are vaccinated and those who are not? OurWordInDataOrg 2021. https://ourworldindata.org/covid-deaths-by-vaccination (accessed February 6, 2022).

[4]　Noor R, Shareen S, Billah M. COVID-19 vaccines: their effectiveness against the severe acute respiratory syndrome coronavirus 2 (SARS-CoV-2) and its emerging variants. Bulletin of the National Research Centre 2022 46:1 2022;46:1–12. https://doi.org/10.1186/S42269-022-00787-Z.

[5]　Zhou L, Puthenkalam JJ, Zhou L, Puthenkalam JJ. Effects of the Human Development Index on COVID-19 Mortality Rates in High-Income Countries. European Journal of Development Studies 2022;2:26–31. https://doi.org/10.24018/EJDEVELOP.2022.2.3.104.

[6]　Jabłońska K, Aalléa S, Toumi M. The real-life impact of vaccination on COVID-19 mortality in Europe and Israel. Public Health 2021;198:230–7. https://doi.org/10.1016/J.PUHE.2021.07.037.

[7]　Papadopoulos VP, Emmanouilidou A, Yerou M, Panagaris S, Souleiman C, Varela D, et al. SARS-CoV-2 Vaccination Coverage and Key Public Health Indicators May Explain Disparities in COVID-19 Country-

[ ] Specific Case Fatality Rate Within European Economic Area. Cureus 2022;14. https://doi.org/10.7759/CUREUS.22989.

[8] Ziakas PD, Kourbeti IS, Mylonakis E. Comparative Analysis of Mortality From Coronavirus Disease 2019 Across the European Union Countries and the Effects of Vaccine Coverage. Open Forum Infectious Diseases 2022;9. https://doi.org/10.1093/OFID/OFAC006.

[9] Ylli A, Burazeri G, Wu YY, Sentell T. COVID-19 excess death rate in Eastern European countries associated with weaker regulation implementation and lower vaccination coverage. MedRxiv 2022:2022.02.06.22270549. https://doi.org/10.1101/2022.02.06.22270549.

[10] Ning C, Wang H, Wu J, Chen Q, Pei H, Gao H. The COVID-19 Vaccination and Vaccine Inequity Worldwide: An Empirical Study Based on Global Data. International Journal of Environmental Research and Public Health 2022, Vol 19, Page 5267 2022;19:5267. https://doi.org/10.3390/IJERPH19095267.

[11] Liang LL, Kuo HS, Ho HJ, Wu CY. COVID-19 vaccinations are associated with reduced fatality rates: Evidence from cross-county quasi-experiments. Journal of Global Health 2021;11:1–9. https://doi.org/10.7189/JOGH.11.05019.

[12] Huang C, Yang L, Pan J, Xu X, Peng R. Correlation between vaccine coverage and the COVID-19 pandemic throughout the world: Based on real-world data. Journal of Medical Virology 2022;94:2181–7. https://doi.org/10.1002/JMV.27609.

[13] Udoakang A, Oboh M, Henry-Ajala A, Anyigba C, Omoleke S, Amambua-Ngwa A, et al. Low COVID-19 impact in Africa: The multifactorial Nexus. Open Research Africa 2021;4:47. https://doi.org/10.12688/AASOPENRES.13261.1.

[14] Marziali ME, Hogg RS, Oduwole OA, Card KG. Predictors of COVID-19 testing rates: A cross-country comparison. International Journal of Infectious Diseases 2021;104:370–2. https://doi.org/10.1016/J.IJID.2020.12.083.

[15] Ioannidis JPA. Over- and under-estimation of COVID-19 deaths. European Journal of Epidemiology 2021;36:581–8. https://doi.org/10.1007/S10654-021-00787-9/FIGURES/2.

[16] Cao Y, Hiyoshi A, Montgomery S. COVID-19 case-fatality rate and demographic and socioeconomic influencers: worldwide spatial regression analysis based on country-level data. BMJ Open 2020;10:e043560. https://doi.org/10.1136/BMJOPEN-2020-043560.

[17] Liang LL, Tseng CH, Ho HJ, Wu CY. Covid-19 mortality is negatively associated with test number and government effectiveness. Scientific Reports 2020 10:1 2020;10:1–7. https://doi.org/10.1038/s41598-020-68862-x.

[18] Wei C, Lee CC, Hsu TC, Hsu WT, Chan CC, Chen SC, et al. Correlation of population mortality of COVID-19 and testing coverage: a comparison among 36 OECD countries. Epidemiology & Infection 2021;149. https://doi.org/10.1017/S0950268820003076.

[19] Hasell J, Mathieu E, Beltekian D, Macdonald B, Giattino C, Ortiz-Ospina E, et al. Per capita: COVID-19 tests vs. Confirmed deaths n.d. https://ourworldindata.org/grapher/covid-19-tests-deaths-scatter-with-comparisons (accessed April 19, 2022).

[20] Hasell J, Mathieu E, Beltekian D, Macdonald B, Giattino C, Ortiz-Ospina E, et al. A cross-country database of COVID-19 testing. Scientific Data 2020;7. https://doi.org/10.1038/S41597-020-00688-8.

[21] Hasell J, Mathieu E, Beltekian D, Macdonald B, Giattino C, Ortiz-Ospina E, et al. Coronavirus (COVID-19) Testing - Our World in Data. Our World in Data 2022. https://ourworldindata.org/coronavirus-testing (accessed April 19, 2022).

[22] Iwata K, Miyakoshi C. Is COVID-19 mortality associated with test number? Journal of Family Medicine and Primary Care 2022;11:1842. https://doi.org/10.4103/JFMPC.JFMPC_1633_21.

[23] Mesas AE, Cavero-Redondo I, Álvarez-Bueno C, Cabrera MAS, de Andrade SM, Sequí-Dominguez I, et al. Predictors of in-hospital COVID-19 mortality: A comprehensive systematic review and meta-analysis exploring differences by age, sex and health conditions. PLOS ONE 2020;15:e0241742. https://doi.org/10.1371/JOURNAL.PONE.0241742.

[24] Hoffmann C, Wolf E. Older age groups and country-specific case fatality rates of COVID-19 in Europe, USA and Canada. Infection 2021;49:111–6. https://doi.org/10.1007/S15010-020-01538-W.

[25] Ritchie H, Mathieu E, Rodés-Guirao L, Appel C, Giattino C, Ortiz-Ospina E, et al. Coronavirus Pandemic (COVID-19). OurWordInDataOrg 2020.

[26] UNDP - United Nations Development Program. Human Development Reports 2020. https://hdr.undp.org/ (accessed April 18, 2022).

[27] Hernán MA, Clayton D, Keiding N. The Simpson's paradox unraveled. International Journal of Epidemiology 2011;40:780–5. https://doi.org/10.1093/IJE/DYR041.

# Supplementary material 1: COVID-19 Pandemic Waves in Countries that used Western Vaccines exclusively

A visual comparison of pandemic waves in the 33 countries that only used Western vaccines indicates that, in countries with higher vaccination rates, the winter wave had a flatter mortality curve compared with the first two waves, unlike the countries with lower vaccination rates. This comparison has several notable exceptions, specifically countries that largely avoided the mortality tolls of the first waves, such as Finland, Denmark, and Norway in Europe, as well as New Zealand, Australia, and South Korea.

Countries are ranked from lowest to highest vaccination rates, as of Oct. 2021.

**Chart source:** Google visualization of Coronavirus (COVID-19) statistics data, available on May 26, 2022, based on the COVID-19 Data Repository by the Center for Systems Science and Engineering (CSSE) at Johns Hopkins University, online: https://github.com/CSSEGISandData/COVID-19

| Country | Vaccination – Average Oct. 2021 | Cases | Deaths |
|---|---|---|---|

| Bulgaria | 22.23 | 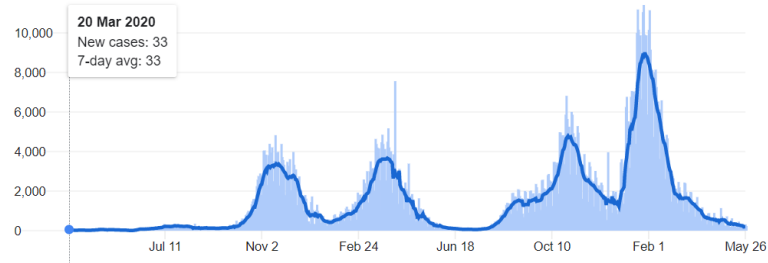 | 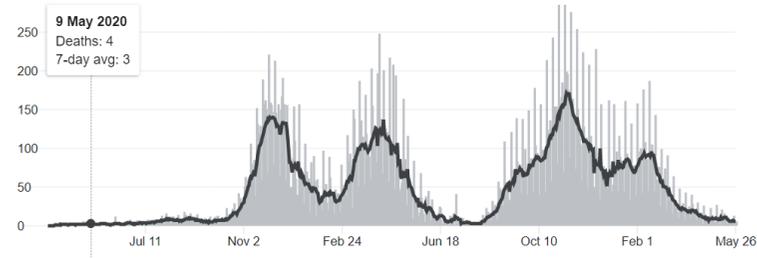 |
| --- | --- | --- | --- |
| Romania | 36.85 | 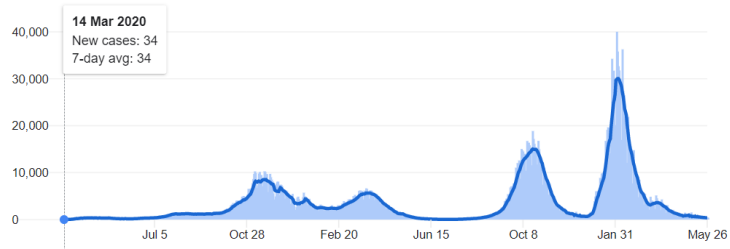 | 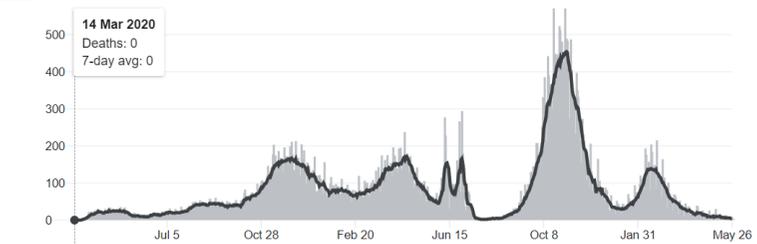 |
| Slovakia | 46.01 | 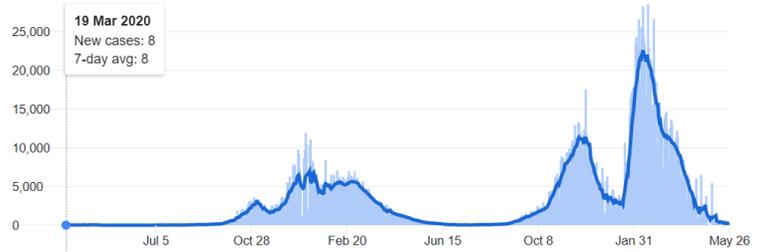 | 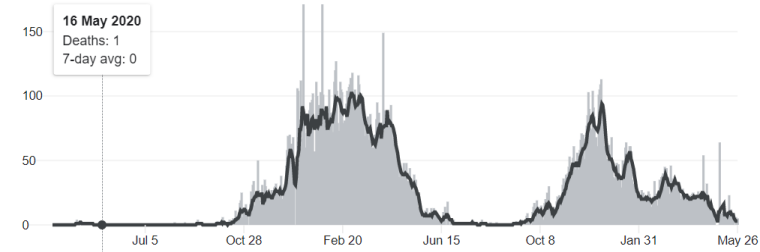 |
| Croatia | 46.84 | 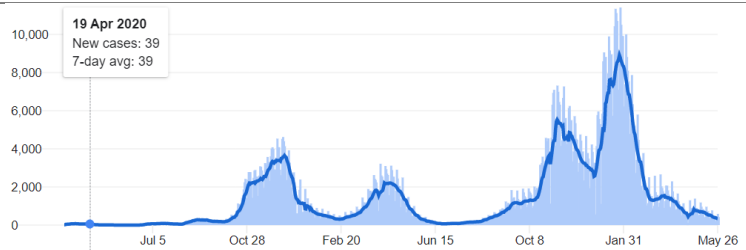 | 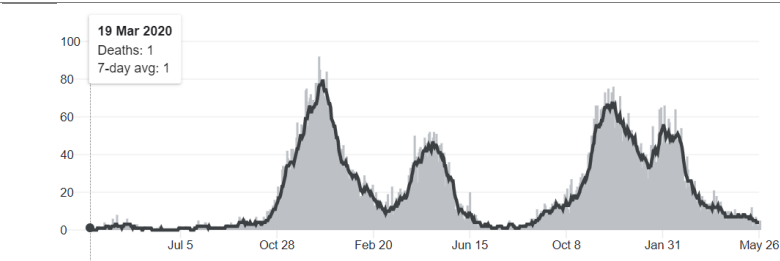 |
| **Poland** | 53.57 | 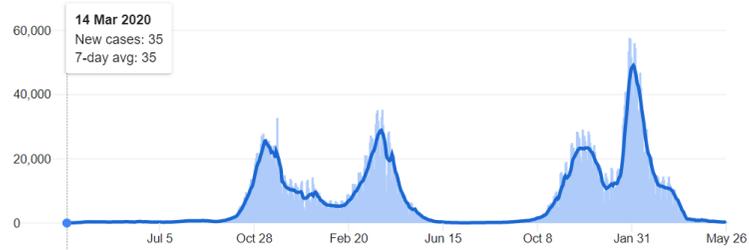 | 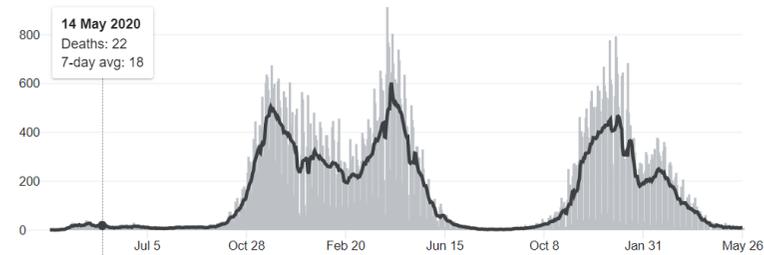 |
| **Slovenia** | 57.19 | 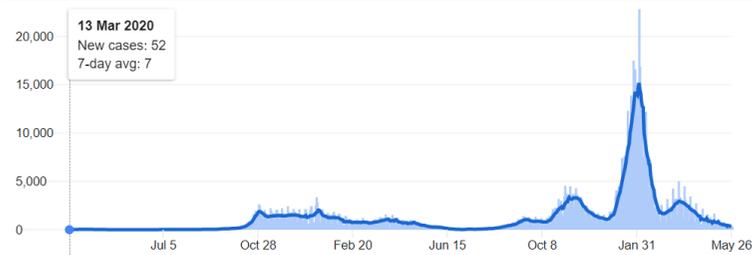 | 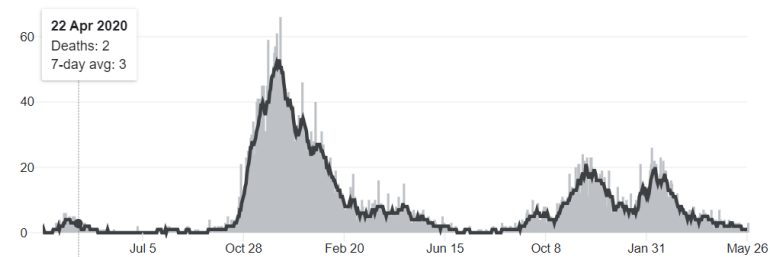 |
| **Czech Republic** | 58.18 | 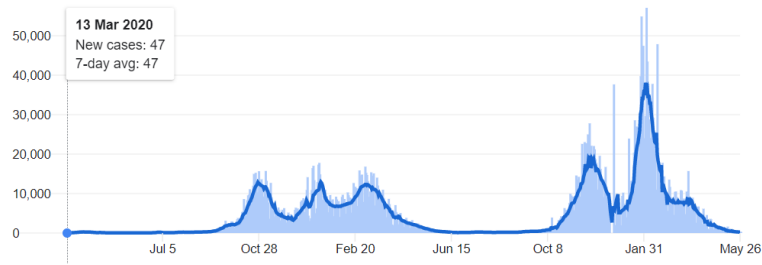 | 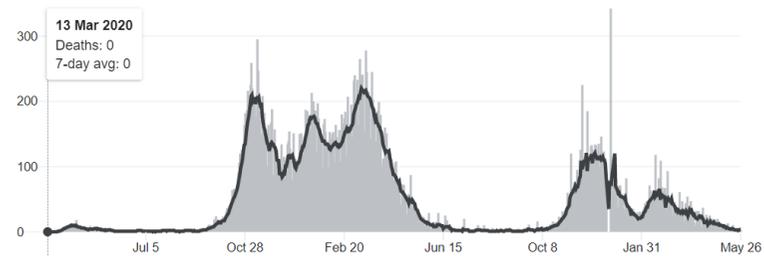 |
| **Estonia** | 60.05 | 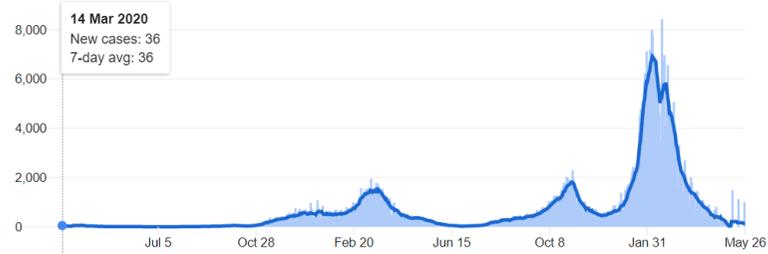 | 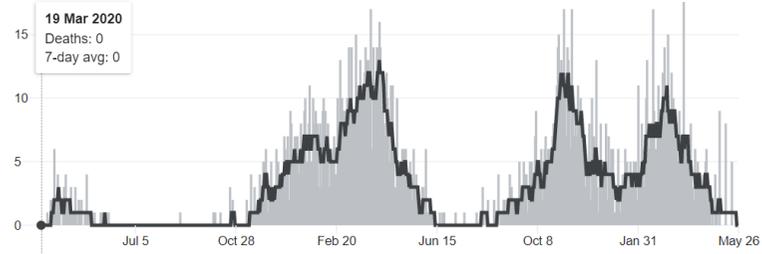 |
| **Latvia** | 62.20 | 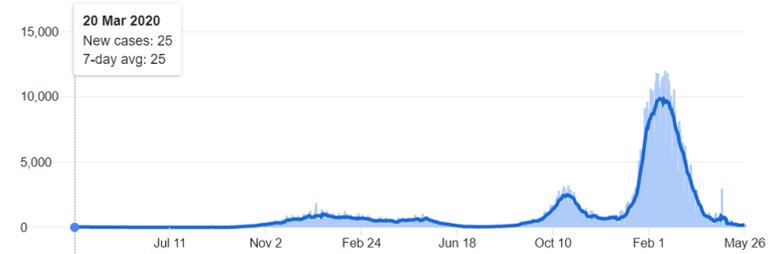 | 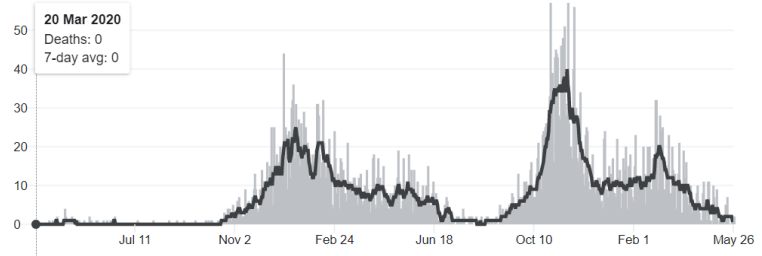 |
| **Greece** | 64.01 | 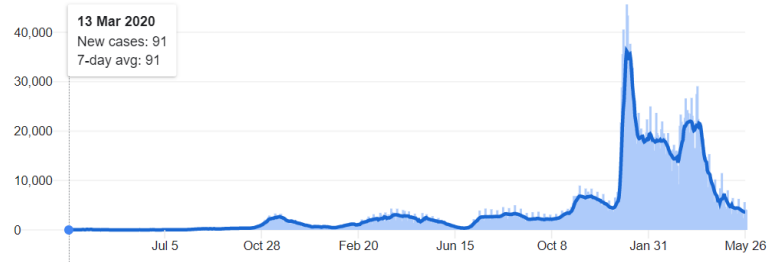 | 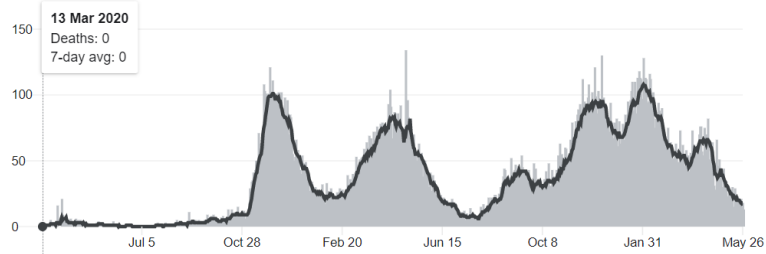 |
| **Austria** | 65.61 | 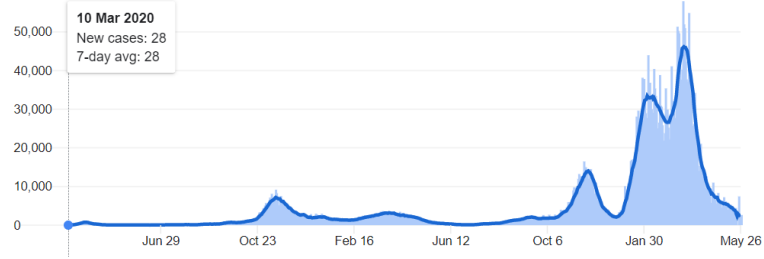 | 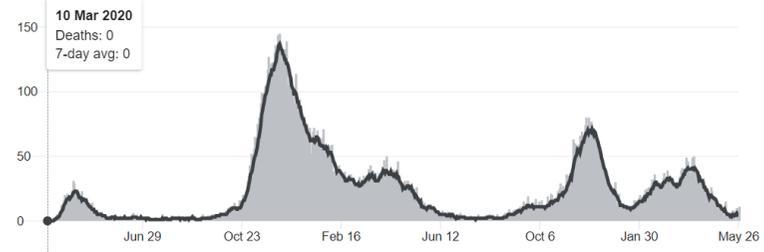 |
| **Switzerland** | 65.93 | 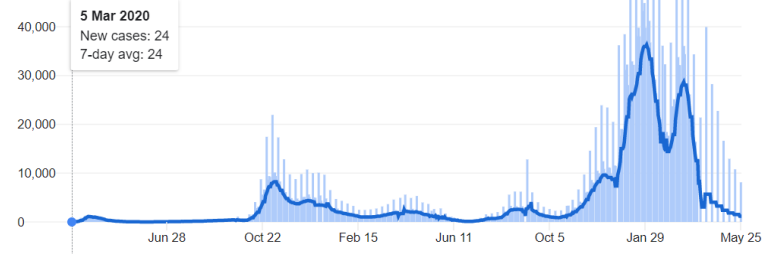 | 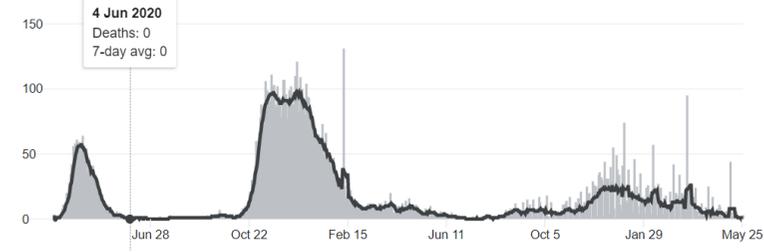 |
| **Israel** | 67.21 | 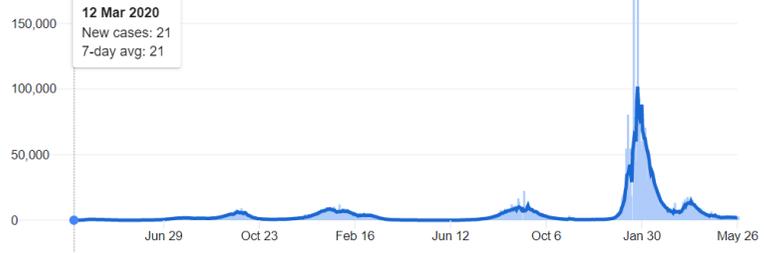 | 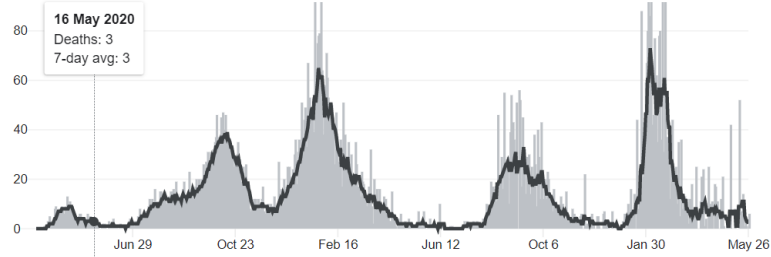 |

| United States | 67.30 | 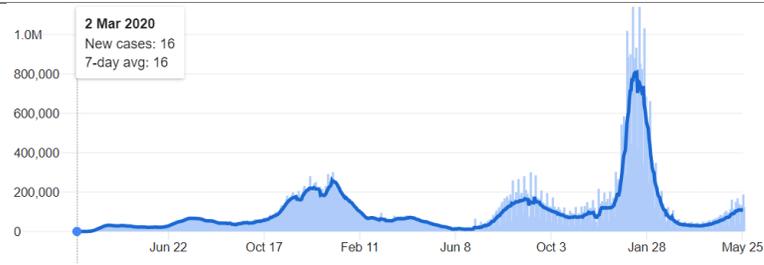 | 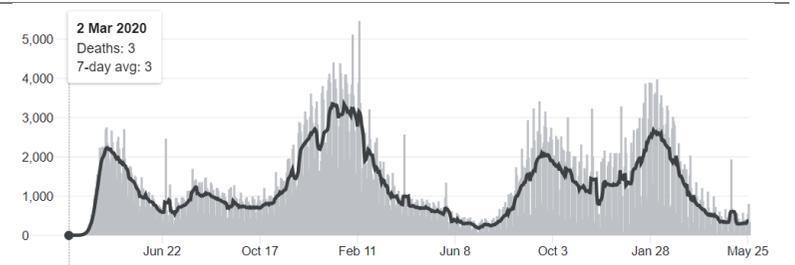 |
| Lithuania | 67.36 | 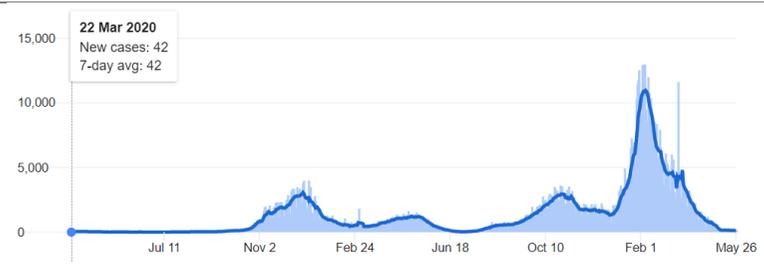 | 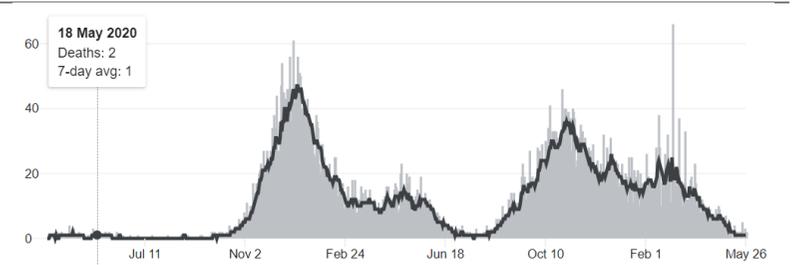 |
| Germany | 68.92 | 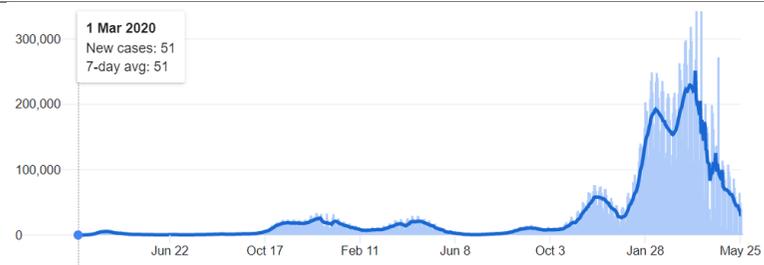 | 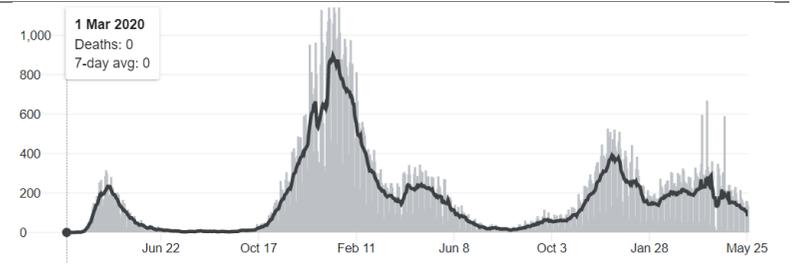 |
| **New Zealand** | 72.39 | 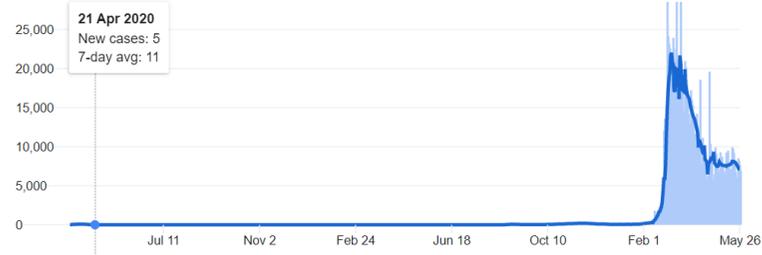 | 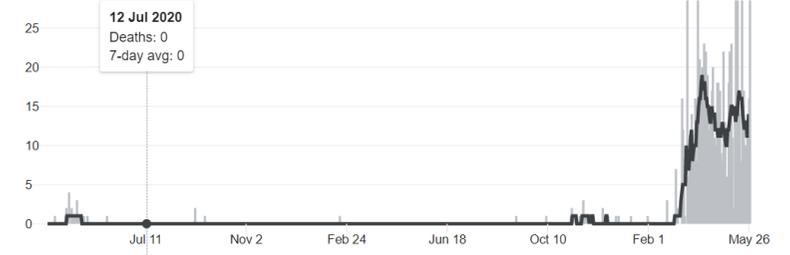 |
| **Sweden** | 73.14 | 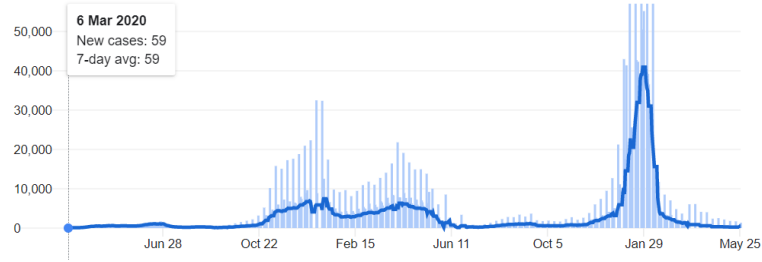 | 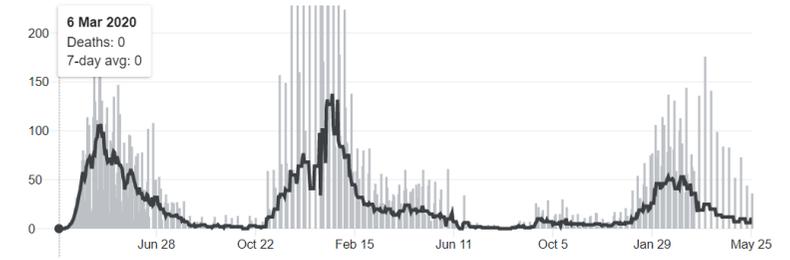 |
| **United Kingdom** | 73.29 | 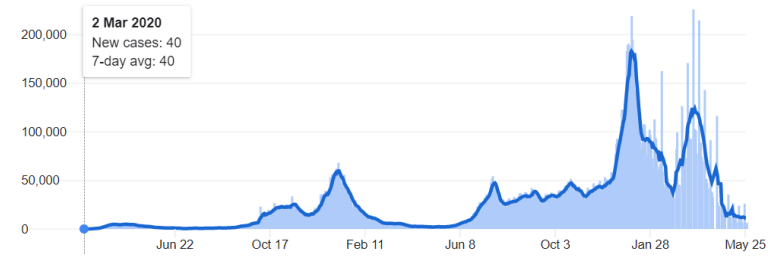 | 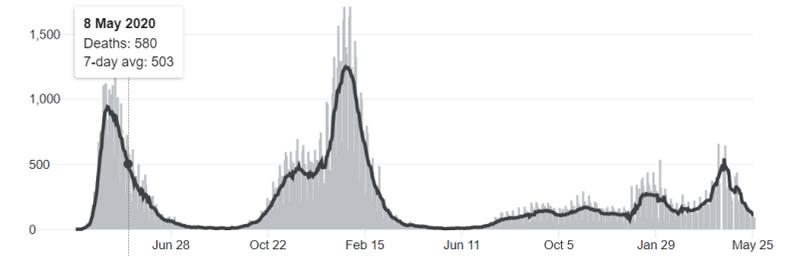 |
| **Australia** | 73.88 | 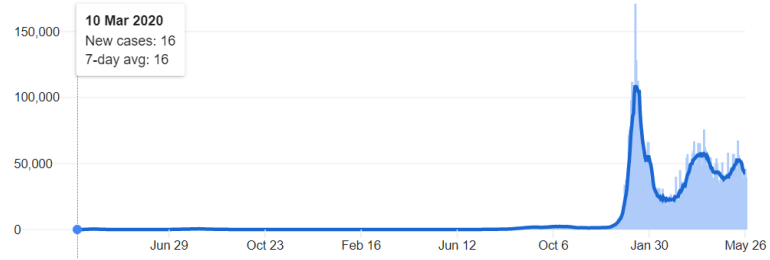 | 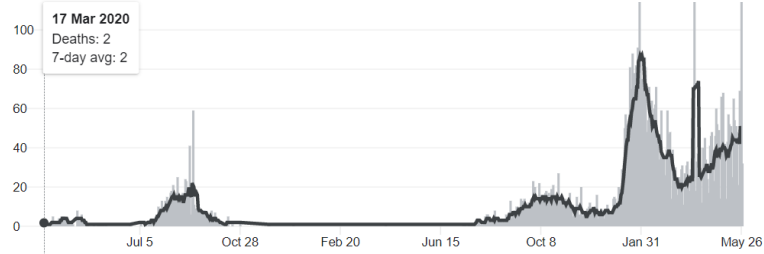 |
| **Belgium** | 75.09 | 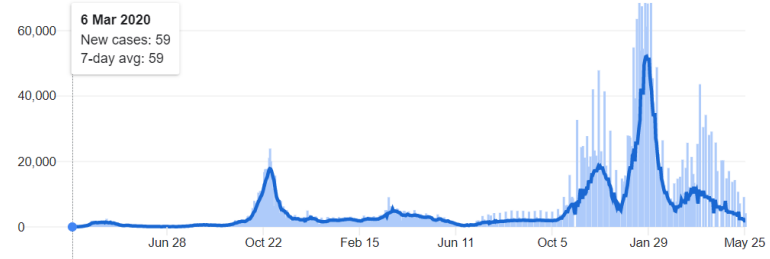 | 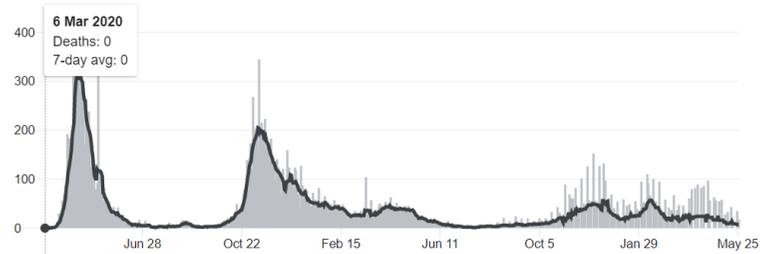 |
| **Netherlands** | 75.87 | 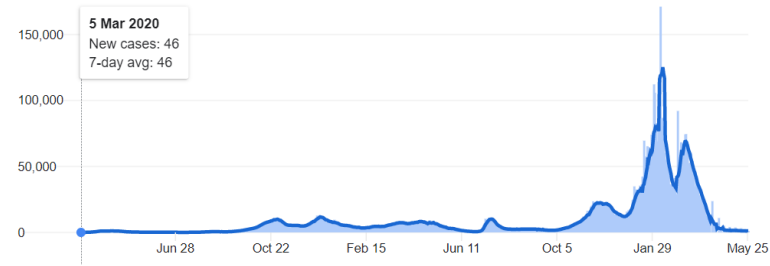 | 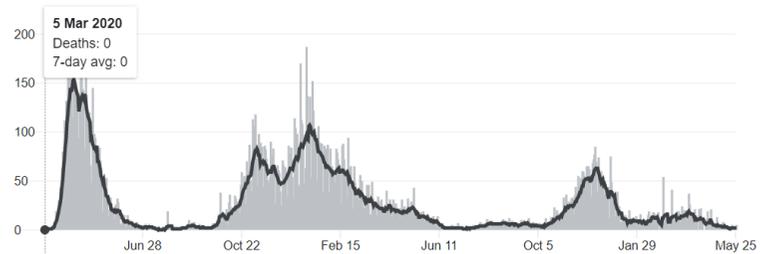 |
| **France** | 76.18 | 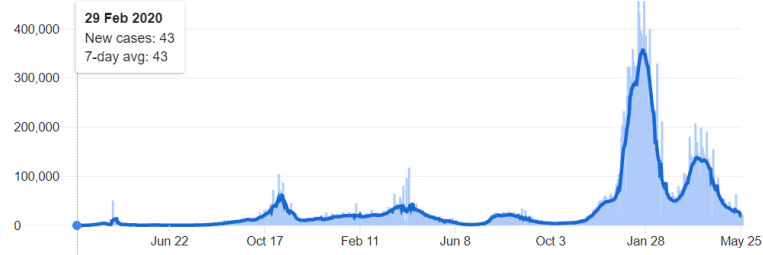 | 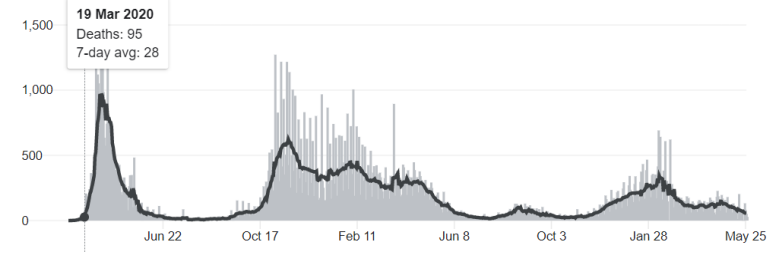 |
| **Finland** | 76.29 | 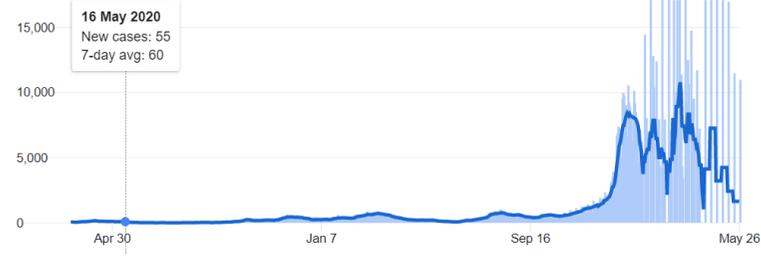 | 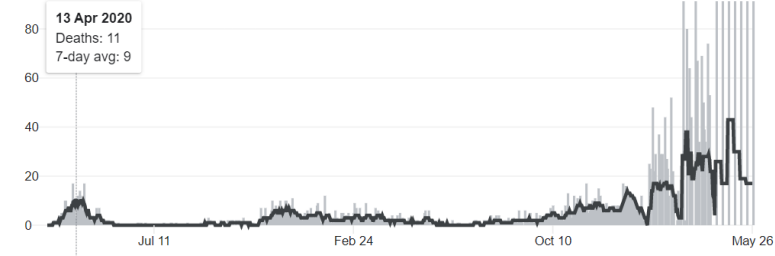 |
| **Ireland** | 76.71 | 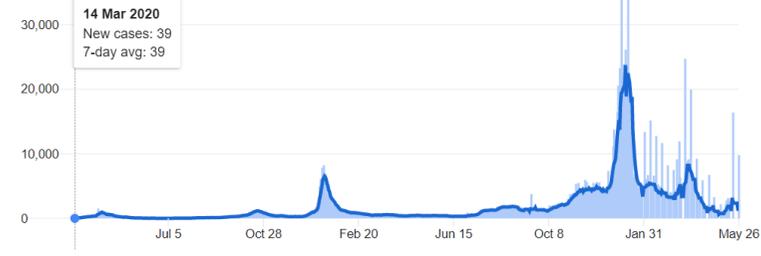 | 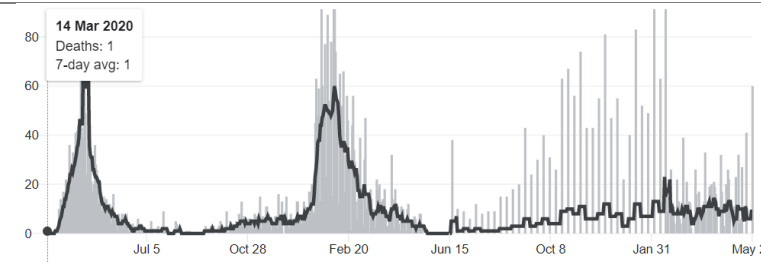 |
| **Denmark** | 77.01 | 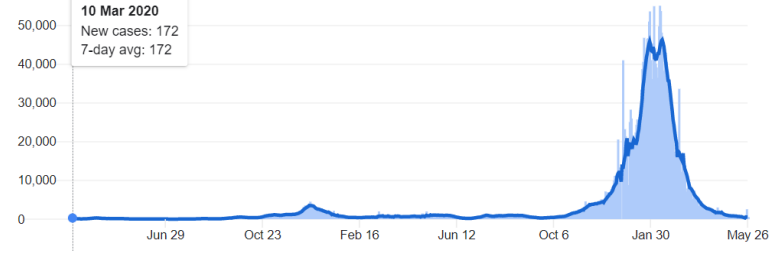 | 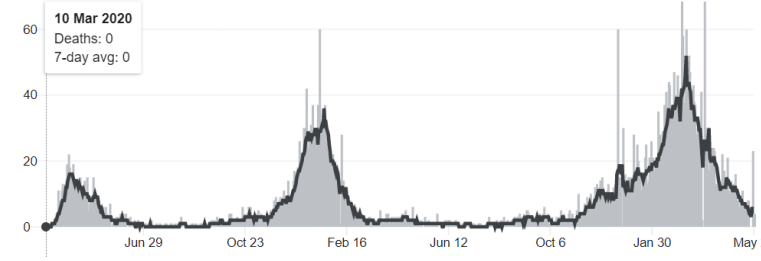 |

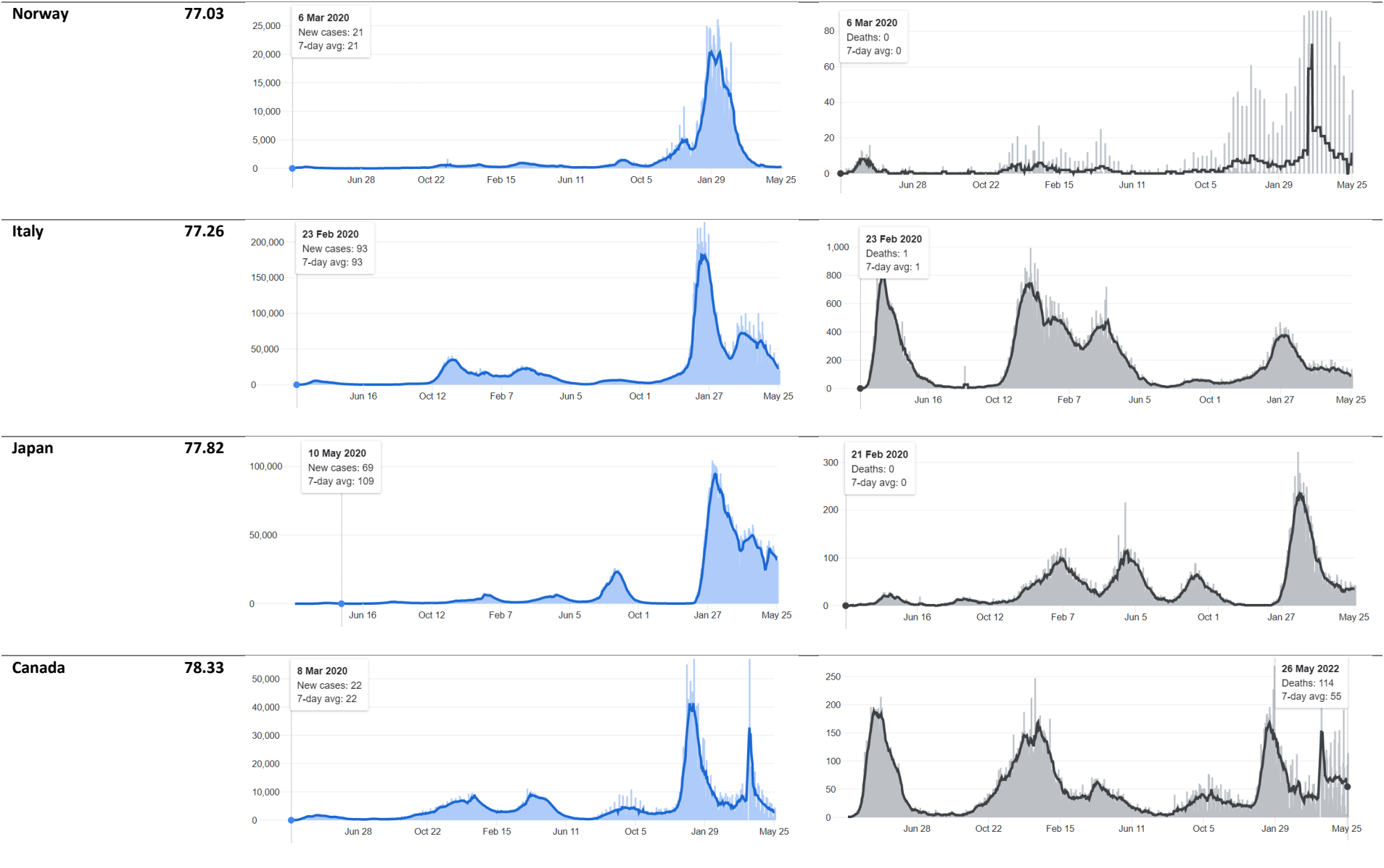

| South Korea | 80.35 | 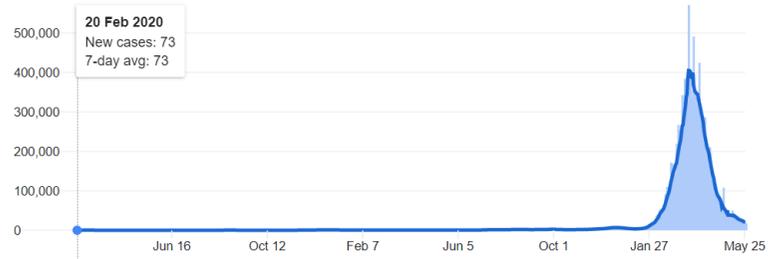 | 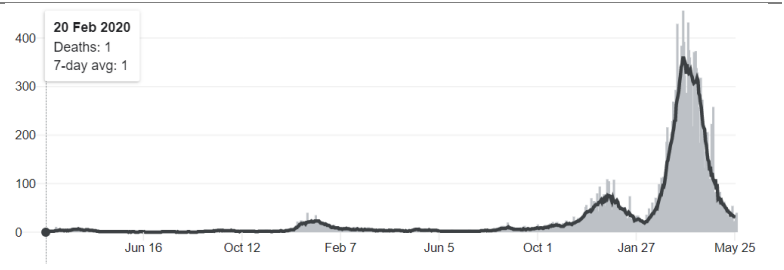 |
| Spain | 81.37 | 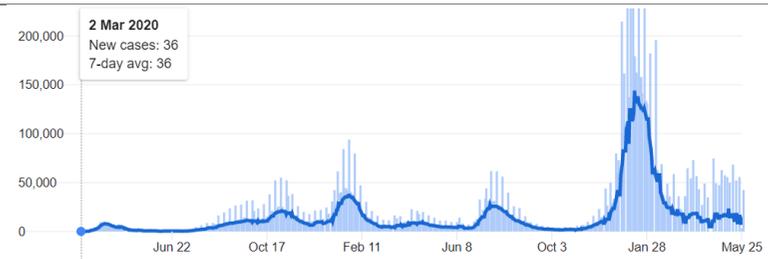 | 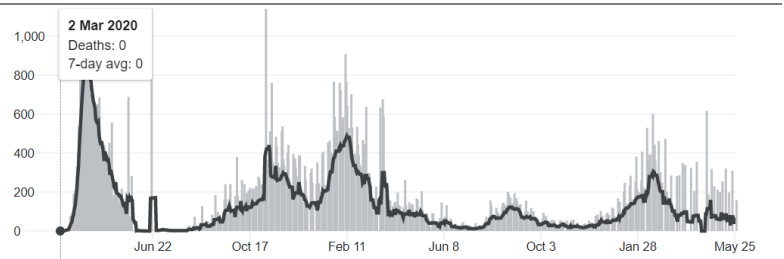 |
| Portugal | 88.69 | 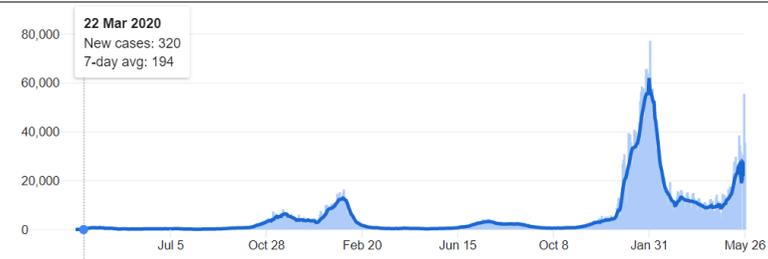 | 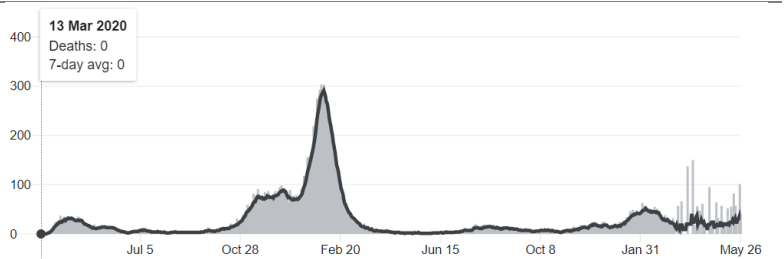 |

# Supplementary material 2

*Table S. M. 1. Descriptive statistics for variables included in analysis*

|  | N | Minimum | Maximum | Mean | Median | Std. Deviation |
|---|---|---|---|---|---|---|
| Cumulative Mortality from Oct. 2021 to March 2022 | 136 | 0 | 1817 | 297.47 | 151.33 | 408.35 |
| Vaccination Rate Oct. 2021 | 135 | 1 | 95 | 41.02 | 39.45 | 27.50 |
| Total Tests per Thousand | 98 | 12 | 13702 | 1478.87 | 758.85 | 2280.59 |
| People aged 65 and older (%) | 134 | 1 | 27 | 9.41 | 6.92 | 6.53 |
| GNI per capita 2019 | 133 | 1035 | 88155 | 20510.44 | 13663.61 | 19373.54 |
| Mean years of schooling 2019 | 135 | 2 | 14 | 9.03 | 9.56 | 3.13 |
| Valid N (listwise) | 96 | | | | | |

*Table S. M. 2. Correlation table for variables included in analysis, linear scale*

|  |  | Cumulative Mortality from Oct. 2021 to March 2022 | Vaccination Rate Oct. 2021 | Total Tests per Thousand | People aged 65 and older (%) | GNI per capita 2019 | Mean years of schooling 2019 |
|---|---|---|---|---|---|---|---|
| Cumulative Mortality from Oct. 2021 to March 2022 | Pearson Correlation | 1 | .241** | .281** | .648** | .335** | .544** |
|  | Sig. (2-tailed) |  | .005 | .005 | .000 | .000 | .000 |
|  | N | 136 | 135 | 98 | 134 | 133 | 135 |
| Vaccination Rate Oct. 2021 | Pearson Correlation | .241** | 1 | .425** | .626** | .746** | .688** |
|  | Sig. (2-tailed) | .005 |  | .000 | .000 | .000 | .000 |
|  | N | 135 | 135 | 98 | 133 | 132 | 134 |
| Total Tests per Thousand | Pearson Correlation | .281** | .425** | 1 | .337** | .585** | .458** |
|  | Sig. (2-tailed) | .005 | .000 |  | .001 | .000 | .000 |
|  | N | 98 | 98 | 98 | 97 | 96 | 98 |
| People aged 65 and older (%) | Pearson Correlation | .648** | .626** | .337** | 1 | .700** | .747** |
|  | Sig. (2-tailed) | .000 | .000 | .001 |  | .000 | .000 |
|  | N | 134 | 133 | 97 | 134 | 132 | 133 |
| GNI per capita 2019 | Pearson Correlation | .335** | .746** | .585** | .700** | 1 | .742** |
|  | Sig. (2-tailed) | .000 | .000 | .000 | .000 |  | .000 |
|  | N | 133 | 132 | 96 | 132 | 133 | 133 |
| Mean years of schooling 2019 | Pearson Correlation | .544** | .688** | .458** | .747** | .742** | 1 |
|  | Sig. (2-tailed) | .000 | .000 | .000 | .000 | .000 |  |
|  | N | 135 | 134 | 98 | 133 | 133 | 135 |

**. Correlation is significant at the 0.01 level (2-tailed).

*Table S. M. 3. Correlation table for variables included in analysis with logarithmic scale for mortality, age structure, and testing intensity*

|  |  | Cumulative mortality March 2022 - Oct. 2021 (Logarithmic Scale) | Vaccination Rate Oct. 2021 | Total Tests per Thousand (Logarithmic Scale) | People aged 65 and older (%) (Logarithmic scale) | GNI per capita 2019 | Mean years of schooling 2019 |
|---|---|---|---|---|---|---|---|
| Cumulative mortality March 2022 - Oct. | Pearson Correlation | 1 | .525** | .684** | .673** | .473** | .652** |
| 2021 (Logarithmic Scale) | Sig. (2-tailed) | | | .000 | .000 | .000 | .000 | .000 |
| | N | | 135 | 134 | 98 | 133 | 132 | 134 |
| Vaccination Rate Oct. 2021 | Pearson Correlation | .525** | 1 | .699** | .644** | .746** | .688** |
| | Sig. (2-tailed) | .000 | | .000 | .000 | .000 | .000 |
| | N | 134 | 135 | 98 | 133 | 132 | 134 |
| Total Tests per Thousand (Logarithmic Scale) | Pearson Correlation | .684** | .699** | 1 | .605** | .751** | .783** |
| | Sig. (2-tailed) | .000 | .000 | | .000 | .000 | .000 |
| | N | 98 | 98 | 98 | 97 | 96 | 98 |
| People aged 65 and older (%) (Logarithmic scale) | Pearson Correlation | .673** | .644** | .605** | 1 | .641** | .758** |
| | Sig. (2-tailed) | .000 | .000 | .000 | | .000 | .000 |
| | N | 133 | 133 | 97 | 134 | 132 | 133 |
| GNI per capita 2019 (Logarithmic scale) | Pearson Correlation | .643** | .807** | .839** | .756** | 1 | .866** |
| | Sig. (2-tailed) | .000 | .000 | .000 | .000 | | .000 |
| | N | 132 | 132 | 96 | 132 | 133 | 133 |
| Mean years of schooling 2019 | Pearson Correlation | .652** | .688** | .783** | .758** | .742** | 1 |
| | Sig. (2-tailed) | .000 | .000 | .000 | .000 | .000 | |
| | N | 134 | 134 | 98 | 133 | 133 | 135 |

**. Correlation is significant at the 0.01 level (2-tailed).

*Table S. M. 4. Country classification on testing rate (tests per thousand – quartiles). Source: Authors' analysis on OWID data*

| Testing Rate Classification | Countries |
|---|---|
| **Countries below the median (N = 49):** | Albania, Angola, Argentina, Azerbaijan, Bangladesh, Bolivia, Bosnia and Herzegovina, Brazil, Colombia, Costa Rica, Dominican Republic, Ecuador, Egypt, Ethiopia, Gabon, Ghana, Guatemala, India, Indonesia, Iran, Iraq, Jamaica, Japan, Madagascar, Mauritania, Mexico, Morocco, Mozambique, Myanmar, Namibia, Nepal, Nigeria, Pakistan, Paraguay, Peru, Philippines, Poland, Rwanda, Senegal, South Africa, South Sudan, Sri Lanka, Taiwan, Thailand, Togo, Trinidad and Tobago, Uganda, Ukraine, Zimbabwe |
| **Countries at and above the median (N = 49):** | Armenia, Australia, Austria, Bahrain, Belarus, Belgium, Botswana, Bulgaria, Canada, Chile, Croatia, Czech Republic, Denmark, Estonia, Finland, France, Georgia, Germany, Greece, Hong Kong, Hungary, Ireland, Israel, Italy, Jordan, Latvia, Lithuania, Malaysia, Mongolia, Netherlands, New Zealand, Norway, Panama, Portugal, Romania, Russia, Saudi Arabia, Serbia, Slovakia, Slovenia, Spain, Sweden, Switzerland, Turkey, United Arab Emirates, United Kingdom, United States, Uruguay, Vietnam |
| **Countries with missing information on testing (N = 38):** | Algeria, Benin, Cambodia, Cameroon, China, Congo, Cuba, El Salvador, Eswatini, Gambia, Guinea, Guinea-Bissau, Honduras, Kazakhstan, Kenya, Kyrgyzstan, Lebanon, Liberia, Libya, Malawi, Mali, Moldova, Nicaragua, Niger, Oman, Palestine, Papua New Guinea, Sierra Leone, Singapore, Somalia, South Korea, Sudan, Syria, Tajikistan, Tunisia, Uzbekistan, Venezuela, Yemen |

*Table S. M. 5. Descriptive statistics differentiated per analytical segments defined through testing intensity*

| | Countries without testing information | | | | | Countries with testing below the median | | | | | Countries with testing at or above the median | | | | |
|---|---|---|---|---|---|---|---|---|---|---|---|---|---|---|---|
| | N | Min. | Max. | Mean | Std. Dev. | N | Min. | Max. | Mean | Std. Dev. | N | Min. | Max. | Mean | Std. Dev. |
| Cumulative Mortality from Oct. 2021 to March 2022 | 38 | 0 | 905 | 78.8 | 159.0 | 49 | | 1464 | 189.5 | 319.7 | 49 | 7 | 1817 | 575.0 | 467.9 |
| Vaccination Rate Oct. 2021 | 37 | 1 | 87 | 24.2 | 26.5 | 49 | 1 | 75 | 33.1 | 23.0 | 49 | 16 | 95 | 61.6 | 18.6 |
| Total Tests per Thousand | | | | | | 49 | 12 | 754 | 267.0 | 197.2 | 49 | 764 | 13702 | 2690.8 | 2733.7 |
| People aged 65 and older (%) | 37 | 2 | 15 | 5.3 | 3.4 | 48 | 2 | 27 | 6.9 | 4.8 | 49 | 1 | 23 | 15.0 | 5.9 |
| GNI per capita 2019 | 37 | 1035 | 88155 | 10056.9 | 15637.6 | 48 | 1250 | 42932 | 10794.2 | 8157.7 | 48 | 7433 | 69394 | 38284.6 | 17181.3 |
| Mean years of schooling 2019 | 37 | 2 | 12 | 7.1 | 3.1 | 49 | 3 | 13 | 7.9 | 2.5 | 49 | 8 | 14 | 11.7 | 1.5 |
| Valid N (listwise) | 37 | | | | | 48 | | | | | 48 | | | | |

*Table S. M. 6. Country classification on age structure (proportion of people aged 65 and more). Source: Authors' analysis on OWID data*

| Age Structure Classification | Countries |
|---|---|
| Countries with proportion of people aged 65+ below the median (N = 67) | Algeria, Angola, Azerbaijan, Bahrain, Bangladesh, Benin, Bolivia, Botswana, Cambodia, Cameroon, Congo, Egypt, Eswatini, Ethiopia, Gabon, Gambia, Ghana, Guatemala, Guinea, Guinea-Bissau, Honduras, India, Indonesia, Iran, Iraq, Jordan, Kenya, Kyrgyzstan, Liberia, Libya, Madagascar, Malawi, Malaysia, Mali, Mauritania, Mexico, Mongolia, Morocco, Mozambique, Myanmar, Namibia, Nepal, Nicaragua, Niger, Nigeria, Oman, Pakistan, Palestine, Papua New Guinea, Paraguay, Philippines, Rwanda, Saudi Arabia, Senegal, Sierra Leone, Somalia, South Africa, South Sudan, Sudan, Tajikistan, Togo, Uganda, United Arab Emirates, Uzbekistan, Venezuela, Yemen, Zimbabwe |
| Countries with proportion of people aged 65+ at or above the median (N = 67) | Albania, Argentina, Armenia, Australia, Austria, Belarus, Belgium, Bosnia and Herzegovina, Brazil, Bulgaria, Canada, Chile, China, Colombia, Costa Rica, Croatia, Cuba, Czech Republic, Denmark, Dominican Republic, Ecuador, El Salvador, Estonia, Finland, France, Georgia, Germany, Greece, Hong Kong, Hungary, Ireland, Israel, Italy, Jamaica, Japan, Kazakhstan, Latvia, Lebanon, Lithuania, Moldova, Netherlands, New Zealand, Norway, Panama, Peru, Poland, Portugal, Romania, Russia, Serbia, Singapore, Slovakia, Slovenia, South Korea, Spain, Sri Lanka, Sweden, Switzerland, Thailand, Trinidad and Tobago, Tunisia, Turkey, Ukraine, United Kingdom, United States, Uruguay, Vietnam |

*Table S. M. 7. Descriptive statistics differentiated per analytical segments defined through age structure*

|  | Proportion of people aged 65+ below the median | | | | | Proportion of people age 65+ at or above the median | | | | |
|---|---|---|---|---|---|---|---|---|---|---|
|  | N | Min. | Max. | Mean | Std. Dev. | N | Min. | Max. | Mean | Std. Dev. |
| Cumulative Mortality from Oct. 2021 to March 2022 | 67 | 0 | 335 | 60.68 | 83.69 | 67 |  | 1817 | 542.65 | 462.31 |
| Vaccination Rate Oct. 2021 | 67 | 1 | 95 | 23.90 | 23.04 | 66 | 16 | 88 | 58.64 | 19.36 |
| Total Tests per Thousand | 39 | 12 | 11223 | 732.59 | 1900.83 | 58 | 115 | 13702 | 2002.57 | 2398.71 |
| People aged 65 and older (%) | 67 | 1 | 7 | 3.96 | 1.38 | 67 | 7 | 27 | 14.86 | 4.88 |
| GNI per capita 2019 | 66 | 1035 | 67462 | 8957.18 | 11293.80 | 66 | 7433 | 88155 | 32319.72 | 18791.99 |
| Mean years of schooling 2019 | 66 | 2 | 12 | 6.87 | 2.64 | 67 | 7 | 14 | 11.18 | 1.83 |
| Valid N (listwise) | 39 |  |  |  |  | 57 |  |  |  |  |

*Table S. M. 8. Country classification on vaccine type. Source: Wikipedia List of COVID-19 vaccine authorizations[1]*

| Vaccine Type Classification | Countries |
|---|---|
|  |  |

---

[1] https://en.wikipedia.org/wiki/List_of_COVID-19_vaccine_authorizations#Oxford%E2%80%93AstraZeneca

| | |
|---|---|
| **Countries with Western vaccines only (N = 33):** | Australia, Austria, Belgium, Bulgaria, Canada, Croatia, Czech Republic, Denmark, Estonia, Finland, France, Germany, Greece, Ireland, Israel, Italy, Japan, Latvia, Lithuania, Netherlands, New Zealand, Norway, Poland, Portugal, Romania, Slovakia, Slovenia, South Korea, Spain, Sweden, Switzerland, United Kingdom, United States |
| **Countries that also or exclusively used Non-Western vaccines (N = 103):** | Albania, Algeria, Angola, Argentina, Armenia, Azerbaijan, Bahrain, Bangladesh, Belarus, Benin, Bolivia, Bosnia and Herzegovina, Botswana, Brazil, Cambodia, Cameroon, Chile, China, Colombia, Congo, Costa Rica, Cuba, Dominican Republic, Ecuador, Egypt, El Salvador, Eswatini, Ethiopia, Gabon, Gambia, Georgia, Ghana, Guatemala, Guinea, Guinea-Bissau, Honduras, Hong Kong, Hungary, India, Indonesia, Iran, Iraq, Jamaica, Jordan, Kazakhstan, Kenya, Kyrgyzstan, Lebanon, Liberia, Libya, Madagascar, Malawi, Malaysia, Mali, Mauritania, Mexico, Moldova, Mongolia, Morocco, Mozambique, Myanmar, Namibia, Nepal, Nicaragua, Niger, Nigeria, Oman, Pakistan, Palestine, Panama, Papua New Guinea, Paraguay, Peru, Philippines, Russia, Rwanda, Saudi Arabia, Senegal, Serbia, Sierra Leone, Singapore, Somalia, South Africa, South Sudan, Sri Lanka, Sudan, Syria, Taiwan, Tajikistan, Thailand, Togo, Trinidad and Tobago, Tunisia, Turkey, Uganda, Ukraine, United Arab Emirates, Uruguay, Uzbekistan, Venezuela, Vietnam, Yemen, Zimbabwe |

*Table S. M. 9. Descriptive statistics differentiated per analytical segments defined through vaccine type*

|  | Countries with Western and Non-Western Vaccines | | | | | Countries with Western Vaccines Only | | | | |
|---|---|---|---|---|---|---|---|---|---|---|
|  | N | Min. | Max. | Mean | Std. Dev. | N | Min. | Max. | Mean | Std. Dev. |
| Cumulative Mortality from Oct. 2021 to March 2022 | 103 | 0 | 1684 | 200.75 | 347.28 | 33 | 56 | 1817 | 599.35 | 441.55 |
| Vaccination Rate Oct. 2021 | 102 | 1 | 95 | 32.90 | 25.88 | 33 | 21 | 88 | 66.12 | 13.91 |
| Total Tests per Thousand | 66 | 12 | 11,223 | 806.35 | 1,572.66 | 32 | 222 | 13,702 | 2,865.94 | 2,853.00 |
| People aged 65 and older (%) | 101 | 1 | 19 | 6.42 | 4.15 | 33 | 12 | 27 | 18.58 | 2.93 |
| GNI per capita 2019 | 101 | 1035 | 88,155 | 12,745.79 | 13904.51 | 32 | 23325 | 69394 | 45,017.61 | 12,575.02 |
| Mean years of schooling 2019 | 102 | 2 | 13 | 7.96 | 2.82 | 33 | 9 | 14 | 12.34 | 1.07 |
| Valid N (listwise) | 65 |  |  |  |  | 31 |  |  |  |  |

*Table S. M. 10. Crosstabulation of vaccine type with age structure and with testing intensity (absolute numbers of countries per cell)*

|  |  | Countries by vaccine type | | |
|---|---|---|---|---|
|  |  | Countries with Western and Non-Western Vaccines | Countries with Western Vaccines Only | Total |
| **Countries by age structure** | Proportion of people aged 65+ below the median | 67 | 0 | 67 |
|  | Proportion of people age 65+ at or above the median | 34 | 33 | 67 |
| **Total** |  | 101 | 33 | 134 |
|  | Testing below the median | 47 | 2 | 49 |

| Countries by testing intensity | Testing at and above the median | 19 | 30 | 49 |
|---|---|---|---|---|
| Total | | 66 | 32 | 98 |

*Table S. M. 11. Regression model for segmentation analysis based on testing intensity*

| Countries by Testing | | Unstandardized Coefficients | | Standardized Coefficients | t | Sig. |
|---|---|---|---|---|---|---|
| | | B | Std. Error | Beta | | |
| **Testing below the median** N = 42 Adjusted R Square = 0.711 | (Constant) | -1.594 | 1.090 | | -1.463 | .151 |
| | Vaccination Rate Oct. 2021 | -.001 | .004 | -.021 | -.175 | .862 |
| | Total Tests per Thousand (Logarithmic Scale) | **.731** | .211 | **.408** | 3.464 | .001 |
| | People aged 65 and older (%) (Logarithmic scale) | .516 | .451 | .163 | 1.144 | .259 |
| | Mean years of schooling 2019 | **.118** | .048 | **.365** | 2.457 | .018 |
| | GNI 2019 per capita (Logarithmic scale) | .104 | .382 | .047 | .271 | .788 |
| **Testing at and above the median** N = 42 Adjusted R Square = 0.692 | (Constant) | 3.190 | .973 | | 3.279 | .002 |
| | Vaccination Rate Oct. 2021 | **-.006** | .003 | **-.244** | -2.068 | .045 |
| | Total Tests per Thousand (Logarithmic Scale) | **.445** | .146 | **.282** | 3.050 | .004 |
| | People aged 65 and older (%) (Logarithmic scale) | **1.378** | .158 | **.808** | 8.716 | .000 |
| | Mean years of schooling 2019 | .003 | .038 | .009 | .076 | .940 |
| | GNI 2019 per capita (Logarithmic scale) | **-.719** | .292 | **-.346** | -2.467 | .018 |

*Table S. M. 12. Regression model for segmentation analysis based on population age structure*

| Countries by age structure | | Unstandardized Coefficients | | Standardized Coefficients | t | Sig. |
|---|---|---|---|---|---|---|
| | | B | Std. Error | Beta | | |
| **Proportion of people aged 65+ below the median** <br><br> N = 33 <br> Adjusted R Square = 0.611 | (Constant) | -.801 | 1.070 | | -.748 | .459 |
| | Vaccination Rate Oct. 2021 | -.006 | .005 | -.194 | -1.090 | .284 |
| | Total Tests per Thousand (Logarithmic Scale) | **.528** | .224 | **.479** | 2.362 | .024 |
| | People aged 65 and older (%) (Logarithmic scale) | **2.164** | .444 | **.514** | 4.876 | .000 |
| | Mean years of schooling 2019 | **.147** | .053 | **.498** | 2.785 | .009 |
| | GNI 2019 per capita (Logarithmic scale) | -.296 | .362 | -.172 | -.816 | .420 |
| **Proportion of people age 65+ at or above the median** <br><br> N = 51 <br> Adjusted R Square = 0.413 | (Constant) | 2.229 | 1.067 | | 2.089 | .042 |
| | Vaccination Rate Oct. 2021 | **-.009** | .003 | **-.415** | -2.707 | .009 |
| | Total Tests per Thousand (Logarithmic Scale) | **.450** | .128 | **.478** | 3.513 | .001 |
| | People aged 65 and older (%) (Logarithmic scale) | .919 | .396 | .352 | 2.322 | .024 |
| | Mean years of schooling 2019 | .001 | .041 | .006 | .036 | .971 |
| | GNI 2019 per capita (Logarithmic scale) | -.350 | .376 | -.218 | -.930 | .357 |

*Table S. M. 13. Regression model for segmentation analysis based on vaccine type*

| Countries by vaccine type | Unstandardized Coefficients | Standardized Coefficients | t | Sig. |
|---|---|---|---|---|

|  |  | B | Std. Error | Beta |  |  |
| --- | --- | --- | --- | --- | --- | --- |
| **Countries with all (Western and Non-Western) vaccines** <br><br> N = 59 <br> Adjusted R Square = 0.752 | (Constant) | -.396 | .841 |  | -.471 | .639 |
|  | Vaccination Rate Oct. 2021 | .000 | .003 | -.010 | -.103 | .918 |
|  | Total Tests per Thousand (Logarithmic Scale) | **.354** | .160 | **.261** | 2.211 | .031 |
|  | People aged 65 and older (%) (Logarithmic scale) | **1.549** | .231 | **.515** | 6.698 | .000 |
|  | Mean years of schooling 2019 | **.112** | .039 | **.346** | 2.843 | .006 |
|  | GNI 2019 per capita (Logarithmic scale) | -.178 | .288 | -.086 | -.619 | .539 |
| **Countries with Western vaccines only** <br><br> N = 25 <br> Adjusted R Square = 0.598 | (Constant) | 2.143 | 2.438 |  | .879 | .388 |
|  | Vaccination Rate Oct. 2021 | **-.017** | .004 | **-.683** | -3.908 | .001 |
|  | Total Tests per Thousand (Logarithmic Scale) | **.461** | .125 | **.454** | 3.686 | .001 |
|  | People aged 65 and older (%) (Logarithmic scale) | 1.204 | .670 | .232 | 1.798 | .084 |
|  | Mean years of schooling 2019 | -.024 | .049 | -.074 | -.497 | .623 |
|  | GNI 2019 per capita (Logarithmic scale) | -.234 | .558 | -.082 | -.419 | .679 |